\documentclass[showpacs,showkeys,11pt,
preprint,preprintnumbers,nofootinbib,
groupedaddress,superscriptaddress,amsmath,amssymb]{revtex4}
\usepackage[dvipdfm]{graphicx}
\usepackage[hypertex]{hyperref}
\usepackage{cancel}

\begin{document}
\title{Multi-$\tau$ signatures at the LHC in the two Higgs doublet model}
\preprint{UT-HET 061}
\author{Shinya Kanemura}
\email{kanemu@sci.u-toyama.ac.jp}
\affiliation{Department of Physics, The University of Toyama, Toyama 930-8555, Japan}
\author{Koji Tsumura}
\email{ko2@phys.ntu.edu.tw}
\affiliation{Department of Physics and Center for Theoretical Sciences, National Taiwan University, Taipei 10617, Taiwan}
\author{Hiroshi Yokoya}
\email{hyokoya@hep1.phys.ntu.edu.tw}
\affiliation{Department of Physics and  Center for Theoretical Sciences, National Taiwan University, Taipei 10617, Taiwan}
\affiliation{National Center for Theoretical Sciences, National Taiwan University, Taipei 10617, Taiwan}

\date{\today}

\begin{abstract}
A detailed simulation study is performed for multi-$\tau$ signatures
 at the Large Hadron Collider, which can be used to probe additional Higgs
 bosons with lepton-specific Yukawa interactions.
Such an extended Higgs sector is introduced in some of new physics
 models at the TeV scale.
We here consider the two Higgs doublet model with the Type-X Yukawa
 interaction, where nonstandard Higgs bosons predominantly decay into
 tau leptons.
These extra Higgs bosons can be pair produced via $s$-channel
 gauge boson mediation at hadron colliders;
$q \bar q\to Z^*\to HA$ and $q \bar q' \to {W^{\pm}}^{*}\to HH^{\pm}$ ($AH^{\pm}$),
 where $H$, $A$ and $H^{\pm}$ are CP-even, odd and charged Higgs bosons,
 respectively.
Consequently, multi-$\tau$ originated signals appear in the final state as a
 promising signature of such a model.
We find that the main background can be considerably reduced by
 requiring the high multiplicity of leptons and tau-jets with appropriate
 kinematical cuts in the final state.
Thus, assuming the integrated luminosity of a hundred of inverse fb, the
 excess can be seen in various three- and four-lepton channels. 
With the integrated luminosity of thousands of inverse fb, the determination of 
 the mass as well as ratios of leptonic decay branching ratios of 
 these Higgs bosons would also be possible.
\end{abstract}
\maketitle

\section{Introduction}
The $SU(3)_C \times SU(2)_L \times U(1)_Y$ gauge structure of the
standard model (SM) for elementary particles has been tested
precisely~\cite{Ref:PDG}.
A missing piece is the Higgs boson, which is responsible for electroweak symmetry
breaking and mass generation mechanisms for all SM particles.
It is expected that the Higgs boson will be discovered at the Large Hadron Collider (LHC)
in near future.
%
The LHC is also searching for the evidence of new physics beyond
the SM~\cite{Ref:LHC-SUSY,Ref:LHC-4G}.
The LHC has already clarified the absence of light new colored
particles; e.g., squarks and gluinos in the supersymmetric
theories~\cite{Ref:LHC-SUSY}, or forth generation quarks~\cite{Ref:LHC-4G}.
However, a light particle without strong interactions has
not been ruled out yet by the LHC data because of small production cross sections.

The Higgs sector is totally unknown, since no Higgs boson has been
discovered yet~\cite{Ref:LEP-Higgs,Ref:LHC-Higgs,Ref:TeV-Higgs}.
In the SM, only one scalar iso-doublet field is introduced to
spontaneously break the electroweak gauge symmetry.
However, since there is no reason for the Higgs sector with only one
doublet, there is a possibility of non-minimal Higgs sectors.
There are two important experimental constraints on extended Higgs
sectors; i.e., the flavor changing neutral current (FCNC) and the
electroweak rho parameter.
In the SM, these constraints are automatically satisfied:
FCNC is suppressed by the Glashow-Illiopoulos-Maiani 
mechanism, and the rho parameter is predicted to be unity at the tree
level due to the custodial $SU(2)$ symmetry.
On the other hand, non-minimal Higgs sectors suffer from both of them in
general.
It is known that in the Higgs sector with only doublets, the rho parameter
is predicted to be unity at the tree level, while Higgs models with higher
representations such as those with triplet fields predict the rho
parameter to be different values from unity.
Therefore, two Higgs doublet models (THDMs) would be a simplest viable
extension of the SM.
However, in the THDM the most general Yukawa interaction predicts FCNC
at the tree level, because both the doublet couples to a fermion so that
the mass matrix and the Yukawa matrix cannot be diagonalized
simultaneously.
In order to avoid this, a discrete symmetry may be introduced under which
different properties are assigned to each scalar doublet~\cite{Ref:GW}.
Under this symmetry, each fermion couples with only one 
scalar doublet, and hence there are no FCNC at the tree level even in the THDM. 

There are four types of Yukawa interactions depending on the
$Z_2$-charge assignments; i.e., Type-I, II, X and Y.
Type-II is the most familiar type of Yukawa interactions in the THDM,
which is the Higgs sector of the minimal supersymmetric standard model (MSSM).
Another interesting possibility would be the Type-X THDM, where one Higgs
doublet couples with quarks and the other does with
leptons~\cite{Ref:Barger,Ref:AKTY,Ref:TypeX}.
The Type-X THDM can appear in the Higgs sector of a gauged
extension of the Type-III seesaw model~\cite{Ref:GaugedTypeIII}, the model
of three-loop seesaw with electroweak baryogenesis~\cite{Ref:AKS} and a
model for positron cosmic ray anomaly~\cite{Ref:Hall}.
In the SM-like limit, where only one of the CP-even Higgs bosons couples
to the gauge bosons, the Yukawa couplings of the other Higgs bosons tend
to be lepton-specific.
Since Yukawa coupling constants are proportional to the mass of fermions, 
these extra Higgs bosons predominantly decay into tau leptons
for the wide range of the parameter space~\cite{Ref:AKTY}.

The tau lepton has a relatively short lifetime as compared with the muon. 
It decays into lighter leptons and/or hadrons with neutrinos in the detector. 
The decay products always produce missing energies, which make event
reconstructions rather complicated. 
However, for an energetic tau lepton, the missing momentum from its decay 
tends to be oriented to the same direction of the charged track~\cite{Ref:HRZ}. 
Therefore, the tau lepton momentum can be approximately reconstructed 
by using the collinear approximation~\cite{Ref:HRZ}.
Furthermore, the decay of the tau lepton is correlated by its polarization, 
which can be used to separate leptonic decays from hadronic decays~\cite{Ref:BHM}.

In this paper, we study multi-$\tau$ signatures at the LHC 
in the lepton-specific THDM in the SM-like limit. 
Masses of extra Higgs bosons can be of the order of a hundred GeV under 
the $B_s \to X_s \gamma$ results. 
Then the gluon fusion mechanism for such extra Higgs bosons is suppressed 
at hadron colliders. 
In this case, extra Higgs bosons can be pair produced; $q \bar q\to Z^*\to HA$ 
and $q \bar q'\to {W^{\pm}}^{*}\to HH^{\pm}$ ($AH^{\pm}$), where $H$, $A$ and 
$H^{\pm}$ are CP-even, odd and charged Higgs bosons, respectively.
Produced Higgs bosons mainly decay into tau leptons because the Yukawa 
coupling constant is proportional the fermion mass. 
These characteristic decay modes can be observed in multi-$\tau$ signatures 
at the LHC.
We perform detailed simulation studies for the pair production of the
extra Higgs bosons where they subsequently decay into multi-$\tau$
states.
It is found that the main background can be considerably reduced by
 requiring the high multiplicity of leptons and tau-jets with appropriate
 kinematical cuts in the final state.
Assuming the integrated luminosity of a hundred of inverse fb, the
 excess can be seen in various three- and four-lepton channels.

This paper is organized as follows. 
In Sec.~II, we summarize the Type-X THDM and give basic constraints on
the model. 
The simulation studies of multi-$\tau$ signatures at the LHC in the
lepton-specific THDM are presented in Sec.~III. 
Summary and discussions are given in Sec.~IV. 

\section{The model and constraints}~\label{model}
The Higgs potential of the THDM is defined as~\cite{Ref:HHG,Ref:Djouadi2}
\begin{align}
{\mathcal V}^\text{THDM} 
&= +m_1^2\Phi_1^\dag\Phi_1+m_2^2\Phi_2^\dag\Phi_2
-m_3^2\left(\Phi_1^\dag\Phi_2+\Phi_2^\dag\Phi_1\right)
+\frac{\lambda_1}2(\Phi_1^\dag\Phi_1)^2
+\frac{\lambda_2}2(\Phi_2^\dag\Phi_2)^2\nonumber \\
&\qquad+\lambda_3(\Phi_1^\dag\Phi_1)(\Phi_2^\dag\Phi_2)
+\lambda_4(\Phi_1^\dag\Phi_2)(\Phi_2^\dag\Phi_1)
+\frac{\lambda_5}2\left[(\Phi_1^\dag\Phi_2)^2
+(\Phi_2^\dag\Phi_1)^2\right], \label{Eq:HiggsPot}
\end{align}
where $\Phi_i(i=1,2)$ are the Higgs doublets with hypercharge $Y=1/2$. 
A softly broken $Z_2$ symmetry is imposed in the model to forbid FCNC at
the tree level, under which the Higgs doublets are transformed as
$\Phi_1\to +\Phi_1$ and $\Phi_2 \to -\Phi_2$~\cite{Ref:GW}. 
The soft-breaking parameter $m_3^2$ and the coupling constant
$\lambda_5$ are complex in general. 
We here take them to be real assuming that CP is conserved in the Higgs
sector. 

The Higgs doublets can be written in terms of the component fields as
\begin{align}
\Phi_i=\begin{pmatrix}i\,\omega_i^+\\\frac1{\sqrt2}(v_i+h_i-i\,z_i)
\end{pmatrix},
\end{align}
where the vacuum expectation values (VEVs) $v_1$ and $v_2$ satisfy 
$\sqrt{v_1^2+v_2^2}=v \simeq 246$ GeV and $\tan \beta \equiv v_2/v_1$.
The mass eigenstates are obtained by rotating the component fields as
\begin{align}
\begin{pmatrix}h_1\\h_2\end{pmatrix}=\text{R}(\alpha)
\begin{pmatrix}H\\h\end{pmatrix},\quad
\begin{pmatrix}z_1\\z_2\end{pmatrix}=\text{R}(\beta)
\begin{pmatrix}z\\A\end{pmatrix},\quad
\begin{pmatrix}\omega_1^+\\\omega_2^+\end{pmatrix}=\text{R}(\beta)
\begin{pmatrix}\omega^+\\H^+\end{pmatrix},
\end{align}
where $\omega^\pm$ and $z$ are the Nambu-Goldstone bosons, $h$, $H$, $A$
and $H^\pm$ are respectively two CP-even, one CP-odd and charged Higgs
bosons, and
\begin{align}
\text{R}(\theta)=\begin{pmatrix}\cos\theta&-\sin\theta \\
\sin\theta&\cos\theta\end{pmatrix}.
\end{align}
The eight parameters $m_1^2$--$m_3^2$ and
$\lambda_1$--$\lambda_5$ are replaced by the VEV $v$, the mixing angles
$\alpha$ and $\tan\beta$, the Higgs boson masses
$m_h^{},m_H^{},m_A^{}$ and $m_{H^\pm}^{}$, and the soft $Z_2$ breaking
parameter $M^2=m_3^2/(\cos\beta\sin\beta)$.
The coupling constants of the CP-even Higgs bosons with weak gauge bosons
$h VV$ and $H VV (V=W,Z)$ are proportional to $\sin(\beta-\alpha)$ and
$\cos(\beta-\alpha)$, respectively.
When $\sin(\beta-\alpha) =1$, only $h$ couples to the gauge bosons
while $H$ decouples. 
We call this limit as the SM-like limit where $h$ behaves 
as the SM Higgs boson~\cite{Ref:GunionHaber,Ref:KOSY}. 

Assuming the discrete symmetry (see TABLE~\ref{Tab:type}), there can be four
types of Yukawa interactions in the THDM, i.e., Type-I, II, X and
Y~\cite{Ref:Barger,Ref:AKTY}; 
\begin{align}
{\mathcal L}_\text{yukawa}^\text{THDM} =
&-{\overline Q}_LY_u\widetilde{\Phi}_uu_R^{}
-{\overline Q}_LY_d\Phi_dd_R^{}
-{\overline L}_LY_\ell\Phi_\ell \ell_R^{}+\text{H.c.},
\end{align}
where $\Phi_f$ ($f=u,d$ or $\ell$) is either $\Phi_1$ or $\Phi_2$.
\begin{table}[tb]
\begin{center}
\begin{tabular}{|c||c|c|c|c|c|c|}
\hline & $\Phi_1$ & $\Phi_2$ & $u_R^{}$ & $d_R^{}$ & $\ell_R^{}$ &
 $Q_L$, $L_L$ \\  \hline
Type-I  & $+$ & $-$ & $-$ & $-$ & $-$ & $+$ \\
Type-II & $+$ & $-$ & $-$ & $+$ & $+$ & $+$ \\
Type-X  & $+$ & $-$ & $-$ & $-$ & $+$ & $+$ \\
Type-Y  & $+$ & $-$ & $-$ & $+$ & $-$ & $+$ \\
\hline
\end{tabular}
\end{center}
\caption{Variation in charge assignments of the softly broken $Z_2$ symmetry~\cite{Ref:AKTY}.}
 \label{Tab:type}
\end{table}
In the Type-I THDM, all fermions obtain their masses from the VEV of
$\Phi_2$.
In the Type-II THDM, masses of up-type quarks are generated by $\Phi_2$
while those of down-type quarks and charged leptons are acquired by
$\Phi_1$.
The Higgs sector of the MSSM is a special THDM, whose Higgs potential is determined 
by gauge coupling constants and whose Yukawa interaction is of Type-II~\cite{Ref:HHG}. 
In the Type-X Yukawa interaction, all quarks couple to $\Phi_2$ while charged leptons 
couple to $\Phi_1$. 
Remaining one is referred to the Type-Y. 
\begin{table}[tb]
\begin{center}
\begin{tabular}{|c||c|c|c|c|c|c|c|c|c|}
\hline
& $\xi_h^u$ & $\xi_h^d$ & $\xi_h^\ell$
& $\xi_H^u$ & $\xi_H^d$ & $\xi_H^\ell$
& $\xi_A^u$ & $\xi_A^d$ & $\xi_A^\ell$ \\ \hline
Type-I
& $c_\alpha/s_\beta$ & $c_\alpha/s_\beta$ & $c_\alpha/s_\beta$
& $s_\alpha/s_\beta$ & $s_\alpha/s_\beta$ & $s_\alpha/s_\beta$
& $\cot\beta$ & $-\cot\beta$ & $-\cot\beta$ \\
Type-II
& $c_\alpha/s_\beta$ & $-s_\alpha/c_\beta$ & $-s_\alpha/c_\beta$
& $s_\alpha/s_\beta$ & $c_\alpha/c_\beta$ & $c_\alpha/c_\beta$
& $\cot\beta$ & $\tan\beta$ & $\tan\beta$ \\
Type-X
& $c_\alpha/s_\beta$ & $c_\alpha/s_\beta$ & $-s_\alpha/c_\beta$
& $s_\alpha/s_\beta$ & $s_\alpha/s_\beta$ & $c_\alpha/c_\beta$
& $\cot\beta$ & $-\cot\beta$ & $\tan\beta$ \\
Type-Y
& $c_\alpha/s_\beta$ & $-s_\alpha/c_\beta$ & $c_\alpha/s_\beta$
& $s_\alpha/s_\beta$ & $c_\alpha/c_\beta$ & $s_\alpha/s_\beta$
& $\cot\beta$ & $\tan\beta$ & $-\cot\beta$ \\
\hline
\end{tabular}
\end{center}
\caption{The mixing factors in each type of Yukawa interactions in
 Eq.~\eqref{Eq:Yukawa}~\cite{Ref:AKTY}.} \label{Tab:MixFactor}
\end{table}
The Yukawa interactions are expressed in terms of mass eigenstates of
the Higgs bosons as
\begin{align}
{\mathcal L}_\text{yukawa}^\text{THDM} =
&-\sum_{f=u,d,\ell} \Bigl[
+\frac{m_f}{v}\xi_h^f{\overline f}fh
+\frac{m_f}{v}\xi_H^f{\overline f}fH
-i\frac{m_f}{v}\xi_A^f{\overline f}\gamma_5fA
\Bigr] \nonumber\\
&-\Bigl\{ +\frac{\sqrt2V_{ud}}{v}\overline{u}
\bigl[ +m_u\xi_A^u\text{P}_L+m_d\xi_A^d\text{P}_R\bigr]d\,H^+
+\frac{\sqrt2m_\ell\xi_A^\ell}{v}\overline{\nu_L^{}}\ell_R^{}H^+
+\text{H.c.} \Bigr\},\label{Eq:Yukawa}
\end{align}
where $P_{L(R)}$ are projection operators for left-(right-)handed fermions,
and the factors $\xi^f_\varphi$ are listed in TABLE~\ref{Tab:MixFactor}.

Experimental constraints on masses of Higgs bosons $H$, $A$, $H^\pm$ in 
THDMs depend on the type of the Yukawa interaction. 
%
Masses of neutral bosons have been bounded by LEP experiment to be
$m_H^{} > 92.8$~GeV and $m_A^{}> 93.4$~GeV in the MSSM (Type-II THDM
with additional relations)~\cite{Ref:PDG}. 
At the LHC, neutral Higgs bosons can be produced via gluon fusion $gg\to
\phi^0$~\cite{gf} where we define $\phi^0=H$, $A$, associated production
with heavy quarks $pp\to t\bar t\phi^0$, $b\bar b\phi^0$~\cite{ffH} and
the weak boson mediated processes $pp\to Z^*\to
HA$~\cite{Ref:GunionHaber} and $pp\to {{W^\pm}}^*\to \phi^0
H^\pm$~\cite{AH+}. 
For the large $\tan\beta$ region, stronger mass bounds can be obtained
from these production processes at the Tevatron and the
LHC~\cite{Ref:SUSYHiggsTeV,SUSYHiggsLHC}. 
However, if the Yukawa interactions of $H$ and $A$ are quarkophobic
which is realized in the wide parameter space in the Type-X THDM, these
Higgs bosons are less constrained. 
The search for such Higgs bosons at the LEP experiments is found in
Ref.~\cite{Ref:LEP4tau}. 
%
On the other hand, direct search bounds on charged Higgs boson mass has
also been set by LEP experiments as $m_{H^\pm}^{} >79.3$~GeV by assuming
${\mathcal B}(H^+ \to c\bar s)+{\mathcal B}(H^+ \to
\tau^+\nu)=1$~\cite{Ref:LEP2tau2v}. 
Further stronger bound can be obtained in the Type-II(Y)
THDM~\cite{Ref:Barger,bsg} from the $B_s \to X_s \gamma$ results as
$m_{H^\pm}^{} >295$~GeV~\cite{bsg2}. 
The observation of $B\to \tau \nu$ decay also constrains the mass of
charged Higgs bosons for the large $\tan\beta$ region~\cite{btaunu}. 
The LHC can also search for charged Higgs bosons in various production
processes such as $pp\to H^+H^-$~\cite{H+H-}, $gb\to tH^-$~\cite{gbH+}
and $pp\to W^\pm H^\mp$~\cite{WH+}. 
However, these bounds depend on the types of the Yukawa interactions.
Therefore, the relatively light charged Higgs boson is still allowed by
experimental data in the Type-I(X) THDM. 
In the Type-X THDM, the charged Higgs boson mass can be constrained 
by the leptonic decays of tau leptons~\cite{tauleptonicdecay,tauleptonicdecay2}.
Extra Higgs boson searches in the MSSM (Type-II THDM) have been well 
studied so far in the literature. 
Fermiophobic Higgs scenario in the Type-I THDM has also been 
discussed~\cite{Ref:Fermiphobic}. 

In this paper, we focus on the Higgs boson search in the Type-X THDM, 
which is less constrained by $B$ decay data.
In this model, more than $99\%$ of $H$ and $A$ decay into 
pairs of tau leptons for $\tan\beta\gtrsim 3$ in the SM-like limit; 
$\sin(\beta-\alpha)=1$~\cite{Ref:AKTY}. 
The neutral Higgs bosons would be produced in pair by $q \bar q \to Z^*
\to HA$ process at the LHC. 
These Higgs bosons predominantly decay into a four-$\tau$ state, 
$HA\to (\tau^+\tau^-)(\tau^+\tau^-)$, 
which is the characteristic signal of the Type-X THDM. 
The tau leptons further decay into leptons or hadrons with neutrinos. 
Consequently, there are several {\it four-lepton} final states such as 
$\ell\ell\ell\ell, \ell\ell\ell\tau_h, \ell\ell\tau_h\tau_h,
\ell\tau_h\tau_h\tau_h$ and $\tau_h\tau_h\tau_h\tau_h$ with missing
energies, where $\ell$ denotes an election $e$ or a muon $\mu$, and
$\tau_h$ is hadronic decay products of the tau lepton. 
Although the branching ratios of the $\phi^0\to \mu^+ \mu^-$ decay are not
large; namely ${\mathcal B}(\phi^0\to \mu^+\mu^-)\sim
(m_\mu/m_\tau)^2\sim 0.35\%$, the $HA\to (\mu^+\mu^-)(\tau^+\tau^-)$
decay process would also be a clear signature, since
the invariant mass of the muon pair has a peak at $M_{\mu\mu}\simeq
m_{H}^{}$ and $M_{\mu\mu}\simeq m_{A}^{}$. 
This signature results in $\mu^+\mu^-\ell\ell$, $\mu^+\mu^-\ell\tau_h$
and $\mu^+\mu^-\tau_h\tau_h$ final states.
We will give the results for simulations of these signals in the next
section. 

As for the charged Higgs boson associated production, 
the processes $q \bar q'\to {W^{\pm}}^* \to \phi^0H^\pm$ would be dominant.
For $\tan\beta\gtrsim 2$, more than $99\%$ of charged Higgs bosons
decay into a tau lepton with a neutrino and $0.35\%$ does into a muon 
with a neutrino. 
Therefore, the characteristic signatures for this process consist of 
three tau leptons $\phi^0H^\pm \to (\tau^+\tau^-)(\tau^\pm\nu)$, one tau
lepton with two muons $\phi^0H^\pm \to (\mu^+\mu^-)(\tau^\pm\nu)$ or two
tau leptons with one muon $\phi^0H^\pm \to (\tau^+\tau^-)(\mu^\pm\nu)$. 
These signals result in the {\it three-lepton} final states;
$\ell\ell\ell$, $\ell\ell\tau_h$, $\ell\tau_h\tau_h$ and
$\tau_h\tau_h\tau_h$ with the large missing energy.
In the next section, we study the collider signature of these final states
in details.

\section{Multi-$\tau$ signatures}

In this section, we present the results of our simulation study for 
the multi-$\tau$ signatures at the LHC in order to probe the production of 
neutral and charged Higgs bosons, which predominantly decay into tau
leptons and occasionally into muons.

First, we explain the framework of our simulation and event analysis. 
Second, we present studies for the pair production process of the neutral
Higgs bosons. 
Finally, we present studies for the charged Higgs bosons associated
production with the neutral Higgs bosons. 

\subsection{Framework for event generation and pre-selection}

The signal events are generated by using {\tt
MadGraph/MadEvent}~\cite{Alwall:2011uj}, where the decay of tau leptons
is simulated by using {\tt TAUOLA}~\cite{Jadach:1993hs}.
The partonic events are passed to {\tt PYTHIA}~\cite{Sjostrand:2006za}
for parton showering and hadronization.
Initial-state-radiation (ISR) and final-state-radiation (FSR) effects
are included.
We choose the collision energy to be $14$~TeV, and use the {\tt CTEQ6L} 
parton distribution functions~\cite{Pumplin:2002vw}.
Throughout this paper, we set the masses of extra Higgs bosons to
$m_H^{}=130$ GeV, $m_A^{}=170$ GeV and $m_{H^{\pm}}=150$ GeV, 
and take the SM-like limit; $\sin(\beta-\alpha)=1$.
The total cross section for $pp\to HA$ is estimated to be
$53$~fb at the tree level~\cite{Ref:AKTY}. 
For the charged Higgs boson associated production, $pp\to \phi^0
H^{\pm}$, the total cross sections are estimated to be $125$~fb for
$HH^{\pm}$ production, and $77$~fb for $AH^{\pm}$ production. 
These mass splitting among the extra Higgs bosons are allowed by
electroweak precision data in the SM-like limit~\cite{Ref:THDMEW1,Ref:THDMEW2}. 
Background events for $VV$ ($= ZZ$, $ZW$ and $WW$), $t\bar t$ processes
where the weak bosons decay leptonically and hadronically, and 
$V+$jets ($V=W$ and $Z$) processes followed by leptonic decays of weak
bosons are generated by {\tt PYTHIA}, where the decays of tau leptons
are also handled by {\tt TAUOLA}.
The total cross sections for these processes are given as $108$~pb,
$493$~pb and $110(30)$~nb, respectively for $VV$, $t\bar t$ 
and $W(Z)$+jets production processes by {\tt PYTHIA}. 
We ignore $K$-factor corrections for all the signal and background
processes, for simplicity. \\

First, we perform the pre-selection of the events in various four- and
three-lepton channels.
In order to take into account the detector availability, muons and 
electrons are required to be isolated\footnote{%
The isolation condition for muons is given by
$p_T^\text{cone}/p_T^{\mu} < 0.2$ where $p_T^\text{cone}$ is
the sum of the magnitude of the transverse momentum of the particles
inside the $R=0.4$ cone around the muon.
The isolation condition for electrons is given by
$p_T^{e}/p_T^\text{jet} > 0.95$, where $p_T^\text{jet}$ is the
transverse momentum of the jet which contains the electron itself.
The jet is constructed from the final state hadrons, electrons, photons
and non-isolated muons.
For our isolation conditions, the finding efficiency of muons is
slightly better than that of electrons.}
and have $p_T^{}\ge 15$~GeV and $|\eta|\le2.5$, where $p_T^{}$ is the
transverse momentum, the pseudorapidity is defined as
$\eta=(1/2)\ln[\tan(\theta/2)]$ from the scattering angle $\theta$ in
the laboratory frame. 
Those muons and electrons are counted in the events.
Then, we construct primal jets from the final state hadrons by
anti-$k_T^{}$ algorithm~\cite{Cacciari:2008gp} with $R=0.4$ using
the {\tt FastJet} package~\cite{Cacciari:2005hq}.
Among the constructed primal jets, we identify the tau-jet candidates by
the following criteria; 
\begin{quote}
 a jet with $p_T^{}\ge 10$~GeV and $|\eta|\le2.5$ which contains $1$ or $3$ 
 charged hadrons in a small cone ($R=0.15$) centered at the jet momentum 
 direction with the transverse energy deposit to this small cone more 
 than $95$\% of the jet.
\end{quote}
The $R=0.4$ cone of the primal jet acts as an isolation cone to reduce
the mis-tagging probability for non-tau jets. 
We present an estimation of the tau-tagging efficiency and the mis-tagging
probability in Appendix.~\ref{app:tag}.
The other jets with $p_T^{}\ge25$~GeV and $|\eta|\le5$ are regarded as a hard jet, 
which are used to estimate the hard QCD activity of the event.
Finally, the missing transverse momentum $\vec{\cancel{p}}_T^{}$ is calculated
as a negative vector sum of the transverse momentum of visible particles, 
$\vec{\cancel{p}}_T^{} = -\sum_{\rm vis.}\vec{p}_T^{}$.

We assume that electric charges of tau-jets are measurable.
Using the charges of tau-jets, we require the charge sum of the
four leptons in the four-lepton channels to vanish to reduce the
contributions from the background with mis-identified tau-jets.
After the pre-selection of the signal and background events, we perform
further event analysis for each channel.

\subsection{Neutral Higgs boson pair production}

Here, we present the results of simulation studies for the neutral 
Higgs boson pair production process. 
In the Type-X scenario, this process is characterized by the four-lepton 
signature, where the leptons can be charged leptons $e$, $\mu$ and also
the tau-jet $\tau_h$. 
There are fifteen kinds of the four-lepton channels in total.
For the convenience of our analysis, we divide the leptonic channels
into three categories; the channels which contain two or more muons,
those which contain two or more tau-jets, and those with two or more
electrons. 
However the last channels are difficult to be utilized due to the
limited statistics and the negligibly small branching ratio of the
Higgs bosons into electrons. 

\subsubsection{Four-lepton channels with two or more muons}

As we have mentioned at the end of Sec.~\ref{model}, the dimuon from the
direct decay of the Higgs bosons would be a clear signature in the
Type-X THDM. 
Although the decay branching ratio is only $0.35\%$, there is no way to
ignore this signature. 
First, we explain the study for the $2\mu2\tau_h$ channel in detail, and
then we comment and summarize the other channels with two or more
muons. \\ 

After the pre-selection, the obtained numbers of events for the
$2\mu2\tau_h$ channel are $91$, $441$, $393$ and $15675$ for the $HA$,
$VV$, $t\bar t$ and $V+$jets production processes, respectively,
in our simulation assuming the integrated luminosity of $100$~fb$^{-1}$.
Notice that at the pre-selection, the charges of the muons and the
tau-jets are not required to be opposite in each to collect all the
$2\mu2\tau_h$ signals through the $4\tau$ decay of the Higgs boson
pairs. 
Instead, we require the charge sum of the four leptons to vanish. 
The number of signal events are about two orders of the magnitude
smaller than background contributions. 

\begin{figure}[tb]
 \centering
 \includegraphics[height=7cm]{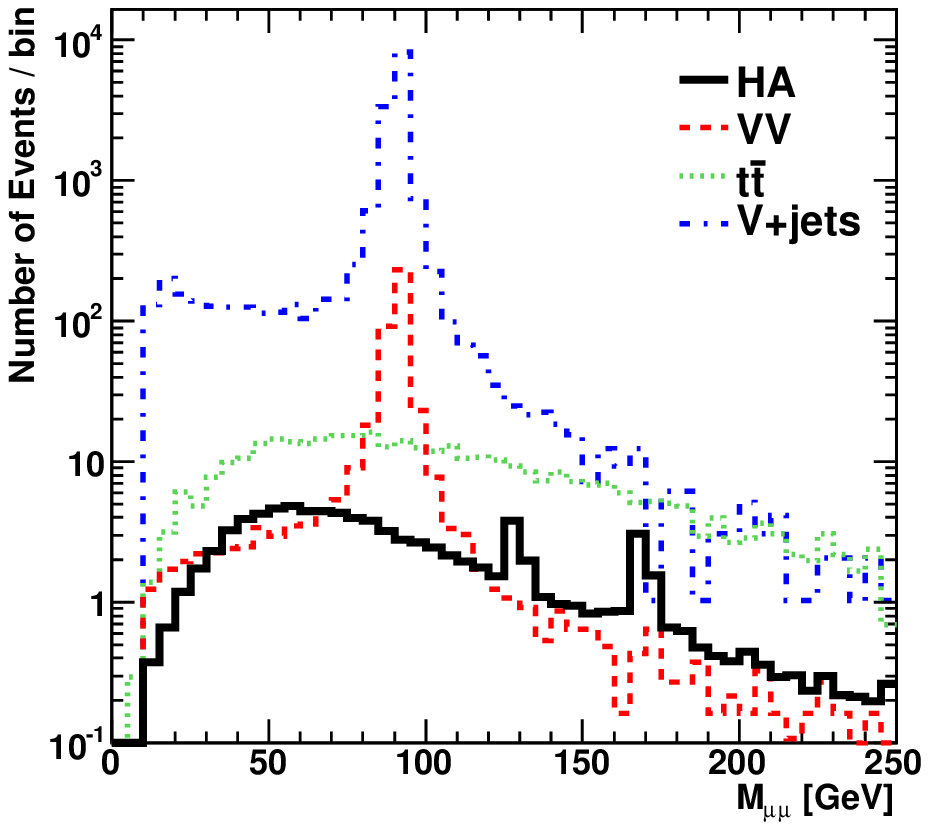}
 \includegraphics[height=7cm]{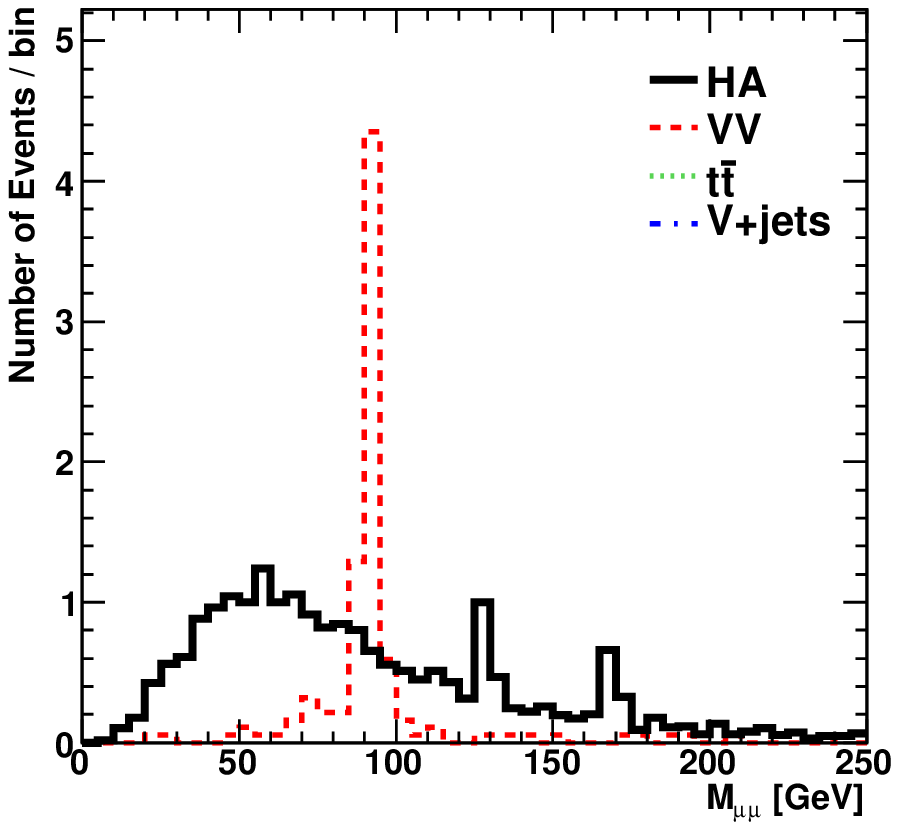}
 \caption{The dimuon invariant mass distribution for the $2\mu2\tau_h$ channel
 after the pre-selection (left) and after the selection cuts before
 the $m_Z^{}$-window cut in TABLE~\ref{Tab:LHC_2T2M} (right).
 The solid black line is for the signal from $HA$ production, the dashed
 red line is for the $VV$ production, the dotted green line is for the
 $t\bar t$ production and the dot-dashed blue line is for the $V+$jets
 production.
 In the signal process, $m_H^{}=130$~GeV and $m_A^{}=170$~GeV are taken.
 }
 \label{FIG:LHC_2T2M_Mmumu}
\end{figure}

In the left panel in FIG.~\ref{FIG:LHC_2T2M_Mmumu}, we show
distributions of the invariant mass $M_{\mu\mu}$ of a muon pair 
in the $2\mu2\tau_h$ channel after the pre-selection.
The solid black line is for the signal $HA$ production, the dashed red line 
is for the $VV$ production, the dotted green line is for the $t\bar t$ production
and the dot-dashed blue line is for the $V+$jets production.
Numbers of events for each process are normalized so as to correspond to 
the integrated luminosity of $100$~fb$^{-1}$.
The invariant mass distribution of the dimuon for the $HA$ production
shows two sharp peaks at $M_{\mu\mu} \simeq 130$~GeV and $170$~GeV which
corresponds to a pair of primary muons from the decay of extra neutral
Higgs bosons.
Due to the secondary muon from the decay of tau leptons, the signal
events behave a broad distribution peaked around $50$~GeV as well.
The dimuon invariant mass distributions for the $VV$ and $V+$jets 
processes have a peak at $M_{\mu\mu}\simeq m_{Z}^{}$ through the decay of 
the $Z$ boson into the dimuon. 
The numbers of events in the distributions for the $VV$ and $V+$jets processes 
decrease rapidly at high mass regions, but that for the $t\bar t$ process 
decreases only slowly. 

\begin{figure}[tb]
 \includegraphics[height=5.3cm]{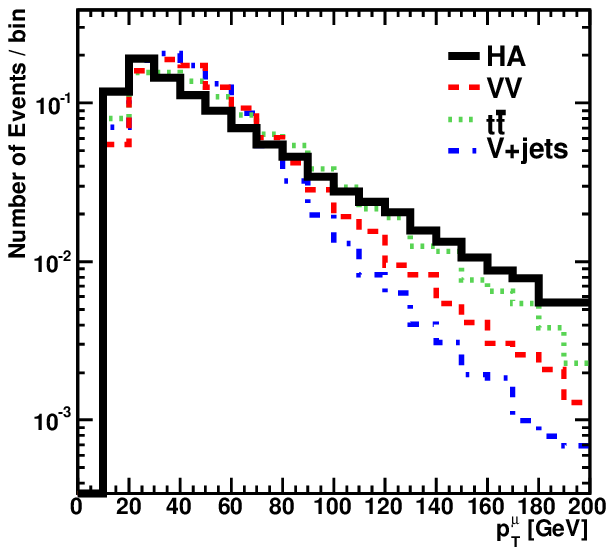}
 \includegraphics[height=5.3cm]{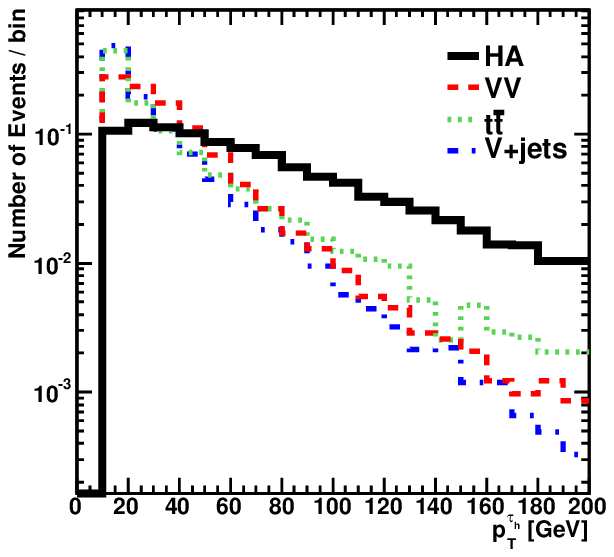}
 \includegraphics[height=5.3cm]{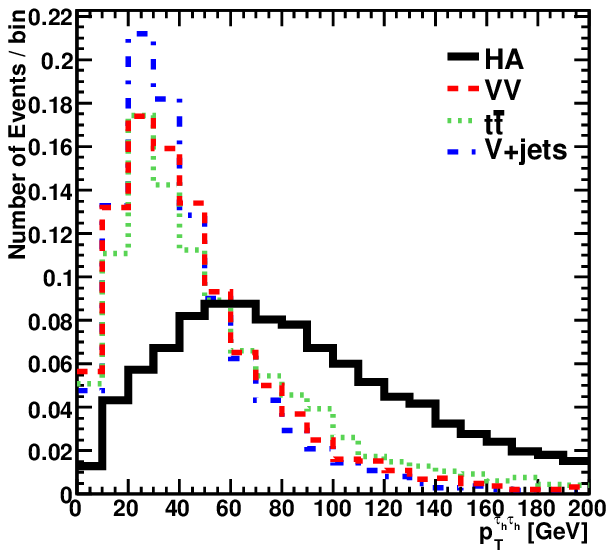}
 \includegraphics[height=5.3cm]{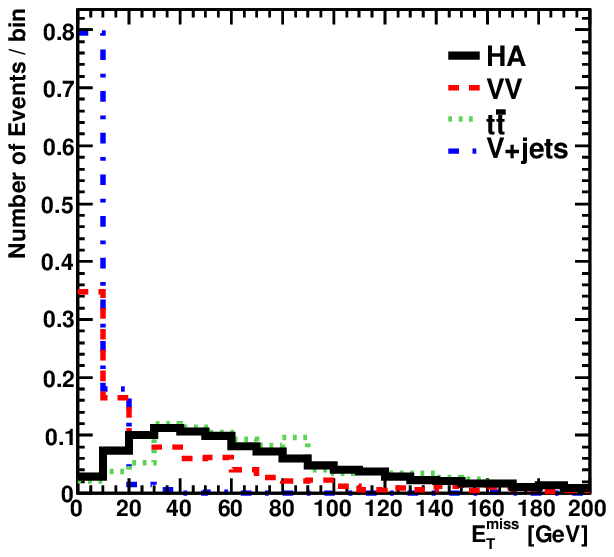}
 \includegraphics[height=5.3cm]{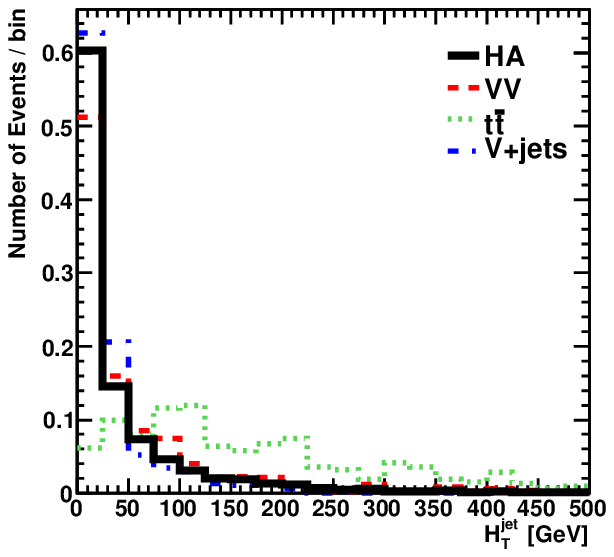}
 \includegraphics[height=5.3cm]{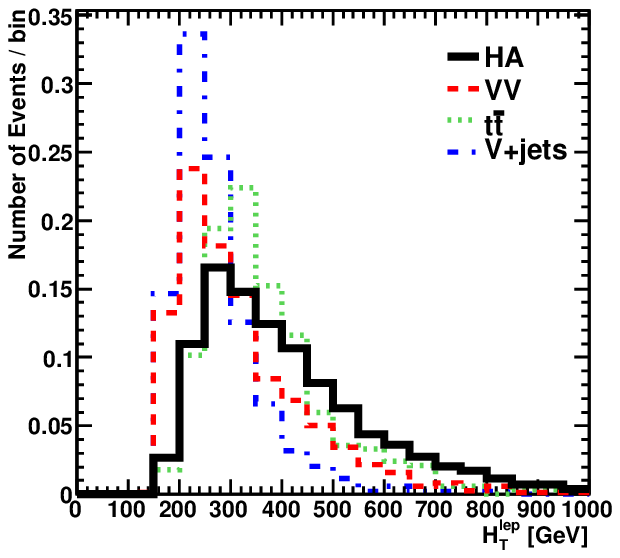}
 \caption{Distributions of kinematical variables for the $2\mu2\tau_h$
 channel.
 Line descriptions follow those in FIG.~\ref{FIG:LHC_2T2M_Mmumu}.}
 \label{FIG:Dist_LHC_2T2M}
\end{figure}
\begin{table}[tb]
 \begin{center}
 \begin{tabular}{|c||c|c||c|c|c||c|c|}
  \hline $2\mu2\tau_h$ event analysis & $HA$ & $\phi^0 H^{\pm}$ &
  $VV$ & $t\bar t$ & $V+$jets & $ s/b $ & $S$ (100~fb$^{-1}$)\\ \hline \hline
  Pre-selection
  & 91.0 & 7.7 & 441. &  393. &  15675.    & $10^{-2}$ &  0.8 \\
  $p_T^{\tau_h} > 40$~GeV
  & 41.1 & 1.7 &  37.7 &  33.4 &   548.   & 0.1 &  1.7 \\
  $\cancel{E}_T^{} > 30$~GeV
  & 33.1 & 1.6 &  14.7 &  29.9 &     5.1   & 0.7 &  4.5 \\
  $H_T^\text{jet} < 50$~GeV
  & 24.0 & 1.2 &   9.7 &   4.3 &     3.1   & 1.4 &  5.1 \\
  $H_T^\text{lep} > 250$~GeV
  & 21.6 & 1.1 &   8.6 &   3.7 &     1.0   & 1.7 &  5.1 \\
  $(m_Z^{})_{\mu\mu} \pm 10$~GeV
  & 18.8 & 1.0 &   2.1 &   3.3 &     1.0   & 3.1 &  5.8 \\
  \hline
  O.S.\ muons
  & 13.2 & 0.6 &   2.0 &   3.3 &     1.0   & 2.2 &  4.4 \\
  $ 0 \le z_{1,2} \le 1.1$
  &  2.9 & 0.1 &   1.0 &   0.6 &     0.    & 1.9 &  1.9 \\
  $(m_Z^{})_{\tau\tau} \pm 20$~GeV
  &  2.7 & 0.1 &   0.2 &   0.6 &     0.    & 3.5 &  2.3 \\
  \hline
  \end{tabular}
 \end{center}
\caption{Table for background reductions in the $2\mu2\tau_h$ channel.
 Listed are the expected number of events for the integrated luminosity
 of 100~fb$^{-1}$ at the LHC $\sqrt{s}=14$~TeV.
 ``O.S.'' stands for ``opposite-sign''.}
\label{Tab:LHC_2T2M}
\end{table}

In FIG.~\ref{FIG:Dist_LHC_2T2M}, we show the distributions of the transverse
momentum $p_T^\mu$ for a muon, that $p_T^{\tau_h}$ for a tau-jet, that 
$p_T^{\tau_h\tau_h}$ for a tau-jet pair, the missing transverse 
energy $\cancel{E}_T(=|\vec{\cancel{p}}_{T}^{}|)$, the hadronic and leptonic 
scalar sum of the transverse momentum $H_T^{\text{jet}} (=
\sum_\text{jet} |p_T^{}|)$ and $H_T^{\text{lep}} (= \sum_{\ell,\tau_h} |p_T^{}|)$,
respectively, in the $2\mu2\tau_h$ channel after the pre-selection. 
In all panels in FIG.~\ref{FIG:Dist_LHC_2T2M}, numbers of the signal and
background events are normalized to be unity. 
Line descriptions follow those in FIG.~\ref{FIG:LHC_2T2M_Mmumu}.
Taking into account the difference of distributions between signal and
background events in FIG.~\ref{FIG:Dist_LHC_2T2M}, we employ the
following selection cuts; 
\begin{subequations}
\begin{align}
 & p_T^{\tau_h}                >  40~\text{GeV}, \label{Cut:2M2T-a} \\
 & \cancel{E}_T^{}             >  30~\text{GeV}, \label{Cut:2M2T-b} \\
 & H_T^\text{jet}              <  50~\text{GeV}, \label{Cut:2M2T-c} \\
 & H_T^\text{lep}              > 250~\text{GeV}, \label{Cut:2M2T-d} \\
 & \left|M_{\mu\mu}-m_Z\right| >  10~\text{GeV}. \label{Cut:2M2T-e}
\end{align}
\end{subequations}
Here, we briefly explain the background reduction strategy with these cuts.
The transverse momentum of tau-jets in signal events is relatively 
larger than that in the background processes due to the heavier mass 
of Higgs bosons than those of $Z$ and $W$ bosons. Therefore a relatively 
high $p_T^{}$ cut for tau-jets could enhance the signal-to-background ratio. 
We note that the high $p_T^{}$ requirement of tau-jets is also suitable 
for the stable tau-tagging efficiency at hadron colliders~\cite{Aad:2011kt}. 
Background events from the $V+$jets process contain two mis-identified
tau-jets from the ISR jets, with a muon pair which comes from
the $Z/\gamma^*\to \mu^+\mu^-$ decay. 
Therefore, $V+$jets background events tend to have small $\cancel{E}_T$, and
the cut on $\cancel{E}_T$ is expected to reduce the $V+$jets background
significantly.
The background contribution from the $t\bar{t}$ events can be reduced by
using the cut on $H_T^\text{jet}$, because the $t\bar t$
events tend to contain many jets due to the $b$ quark fragmentation and
ISR/FSR, even though two of them are mis-identified as tau-jets. 
The cut on $H_T^\text{lep}$ can reduce the $VV$ and $V+$jets backgrounds
significantly. 
Furthermore, the events which contain $Z\to\mu^+\mu^-$ can be reduced 
by rejecting the events with the invariant mass of the muon pair close to
$m_Z^{}$.

The results of the signal/background reduction are summarized 
at each step in TABLE~\ref{Tab:LHC_2T2M}.
We show the expected numbers of events for the integrated luminosity of
$L=100$~fb$^{-1}$ for each process.
In the table, we also include the signal events from the
charged Higgs boson associated production process $pp\to \phi^0 H^\pm$.
The signal-to-background ratio $s/b$ is evaluated at each step of the
cuts, where $s$ and $b$ represent the numbers of signal and background
events, respectively, taking all the $HA$ and $\phi^0 H^{\pm}$
production processes as the signal events. 
In order to evaluate the signal significance, we use the significance
estimator $S$ defined as~\cite{Ref:CMS-TDR} 
\begin{align} 
S &= \sqrt{2 \bigl[ (s+b) \ln(1+s/b)-s \bigr]},
\end{align} 
which is also given at each step of the selection cuts. 
The significance $S$ is proportional to the square root of the
integrated luminosity. 
We choose the selection cuts basically to enhance $S$.
However, since our background events are estimated based on the
leading order cross sections and distributions with limited statistics,
the $s/b$ ratio should not be small but preferably ${\cal O}(1)$ to make
our results conservative.
The largest significance can be obtained after the $m_Z^{}$-window cut,
where the number of the signal events is expected to be about $20$ while 
that of background events is about $6$ giving $s/b\sim 3$ and $S\sim6$
for $L=100$~fb$^{-1}$.

In the right panel in FIG.~\ref{FIG:LHC_2T2M_Mmumu}, we show the dimuon
invariant mass distributions after the selection cuts \eqref{Cut:2M2T-a}, 
\eqref{Cut:2M2T-b}, \eqref{Cut:2M2T-c} and \eqref{Cut:2M2T-d}. 
After these cuts, the signal process dominates the total events 
in the $2\mu2\tau_h$ channel.
Thus the two resonant peaks are more enhanced. 
The mass resolution is expected to be quite well due to the fine
resolution of the muon momentum measurement. 
On the other hand, the expected number of events for $L=100$~fb$^{-1}$
is not sufficient to observe the peaks in the $M_{\mu\mu}$ distribution.

Now let us consider the detail of the events for the signal process from
$HA$ production.
As we have mentioned, the $2\mu2\tau_h$ final state arises through the
$HA\to4\tau$ and $HA\to2\mu2\tau$ decays of the Higgs bosons. 
Assuming the dominant branching ratio of ${\mathcal
B}(\phi^{0}\to\tau^+\tau^-)\simeq1$, the probability to get to the
$2\mu2\tau_h$ final state from the former route is expressed as
${4\choose 2}\,[{\mathcal B}(\tau\to\mu)]^2\,[{\mathcal
B}(\tau\to\tau_h)]^2$, where the binomial coefficient ${4\choose 2}=6$.
On the other hand, the probability through the latter route is
$2\,{\mathcal B}(\phi^0\to\mu^+\mu^-)\,[{\mathcal
B}(\tau\to\tau_h)]^2$.
Thus, the ratio of the probabilities through the two routes is
$3\,[{\mathcal B}(\tau\to\mu)]^2/{\mathcal B}(\phi^0\to\mu^+\mu^-)$,
which is predicted to be about $26$ in the Type-X THDM. 
However, in the actual events, the ratio of the expected numbers of the
events through the two routes suffers from the acceptance cuts on the
muons.
Especially, the $p_{T}^{}$ cut on the muons is expected to reduce the
ratio, since the muons from the decay of tau leptons have relatively
small transverse momentum.
From the left panel in FIG.~\ref{FIG:LHC_2T2M_Mmumu}, the ratio of
the number of the excess events at the sharp $M_{\mu\mu}$ peaks to the
number of the events in the continuous distribution is found to
be $6.6/84.4\sim1./12.8$.
After the selection cuts up to the cut on $H_{T}^\text{lep}$ in
(\ref{Cut:2M2T-d}), the ratio becomes $1.6/20\sim 1./12.5$ (see the
right panel in FIG.~\ref{FIG:LHC_2T2M_Mmumu}). 
Thus, we find our selection cuts do not significantly modify the ratio.
With this fact, the branching ratios of ${\mathcal
B}(\phi^0\to\mu^+\mu^-)$ can be measured as
\begin{align}
 {\mathcal B}(\phi^0\to\mu^+\mu^-) 
 = \frac{\text{Number of the excess events at $M_{\mu\mu}$ peaks}}{\text{Number 
 of events in the continuous dist.}} 
 \times 3\,[{\mathcal B}(\tau\to\mu)]^2 \times \epsilon, \label{Eq:br}
\end{align}
where $\epsilon$ is the correction factor which reflects the
difference of the kinematical distributions of the muons between the
$4\tau$ and $2\mu2\tau$ decays of Higgs bosons resulting the difference
of the kinematical acceptance of the muons.
It could be simulated by using the kinematical distributions of the
muons calculated theoretically and the acceptance cuts for muons. 
In our simulation, we get $\epsilon\sim0.5$, thus the ratio of the muon
acceptance in the $4\tau$ decay to that in the $2\tau2\mu$ decay is
$\sqrt{0.5}\sim 0.7$. 

Once the peaks in the $M_{\mu\mu}$ distribution are observed with the
sufficient number of events, the measurement of the branching ratio into
the dimuon is possible. 
If we consider the case where the Higgs bosons decay into the other
modes as well, such like weak bosons or quarks, the left-hand side of
Eq.~(\ref{Eq:br}) is replaced by the ratio of the branching ratios or
the partial decay widths for muons and tau leptons, which is exactly the
ratio of the square of the Yukawa couplings to muons and tau leptons. \\

Next, we turn our interest to the tau-jet observables.
One of the attractive features of the $2\mu2\tau_h$ channel is that the
four momenta of tau leptons are reconstructable, if the muon pair comes
from the direct decay of one of the neutral Higgs bosons.
In such a case, the missing momentum in the event is expected to come
from the hadronic decay of tau leptons, and then the full kinematics of
tau leptons can be reconstructed by using the collinear approximation
for the relation of the momenta of the tau leptons and the tau-jets.
To enhance the signal events where one of the Higgs bosons decays 
directly into the dimuon, we require the muon pair to have opposite-sign
charges. 
Then, we apply the collinear approximation for the tau-jets to determine 
the four momenta of the tau leptons.

Here, we briefly explain the collinear approximation to calculate the
four momenta of the tau leptons. 
If tau leptons are energetic, the missing momentum from its decay would
be along the direction of the charged track (either a charged hadron
(hadrons) or a charged lepton), ${\vec p}^{\; miss} \simeq c\, {\vec
p}^{\; \tau_j}$, where ${\vec p}^{\; miss}$, ${\vec p}^{\; \tau_j}$ are
the momenta of the neutrino and the charged track, respectively.
The proportionality constant $c$ can be determined by fixing ${\vec
p}^{\; miss}$. 
Accordingly, the momentum of the decaying tau lepton can be
approximately reconstructed as ${\vec p}^{\; \tau} \simeq (1+ c)\, {\vec
p}^{\; \tau_j} \equiv z^{-1}\, {\vec p}^{\; \tau_j}$, where $z$ is the
momentum fraction of the charged track from the parent tau lepton.
At hadron colliders, the transverse components of the missing momentum
$\vec{\cancel{p}}_T^{}$ can be measured.
Assuming that the missing transverse momentum of the event is accounted
solely by the missing particles in the decays of tau leptons, and applying
the collinear approximation for both the tau leptons, the missing transverse
momentum can be expressed by the momenta of charged tracks, as
$\vec{\cancel{p}}_T^{} \simeq {\vec p}^{\; miss_1}_T + {\vec p}^{\;
miss_2}_T \simeq c_1\, {\vec p}^{\; \tau_{j1}}_T + c_2\, {\vec p}^{\;
\tau_{j2}}_T$.
Unknown parameters $c_1$ and $c_2$ are determined by solving simultaneous
equations.
Using the resulting values of $z_1$ and $z_2$, 
the invariant mass of the tau leptons pair is related with that of
the tau-jets pair as $M_{\tau_h\tau_h}^2 \simeq z_1 z_2
M_{\tau\tau}^2$.
The fractions $z_1$ and $z_2$ should be between 0 and 1. 
In our analysis, we set a 10\% margin for the upper bound on the cut 
of the momentum fractions, i.e., $0.<z_{1,2}<1.1$.
This is to take into account the resolution of the momentum
measurements and also the limitation of the validity of the collinear
approximation.
Actually, by taking the 10\% margin the number of the signal events
which have the solution is enhanced by about 10\%, although $s/b$ and
$S$ are almost unchanged. 
The lower limit of $z$ is not significant because of the high $p_T^{}$
cut on the tau-jets.

\begin{figure}[tb]
 \centering
 \includegraphics[height=7cm]{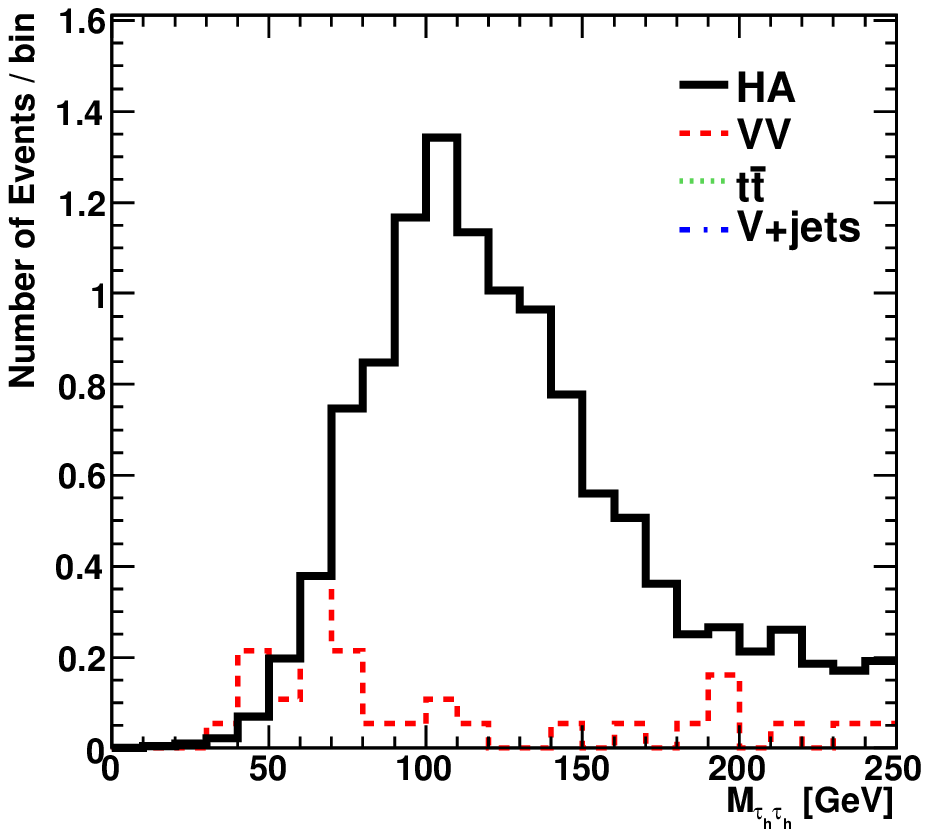}
 \includegraphics[height=7cm]{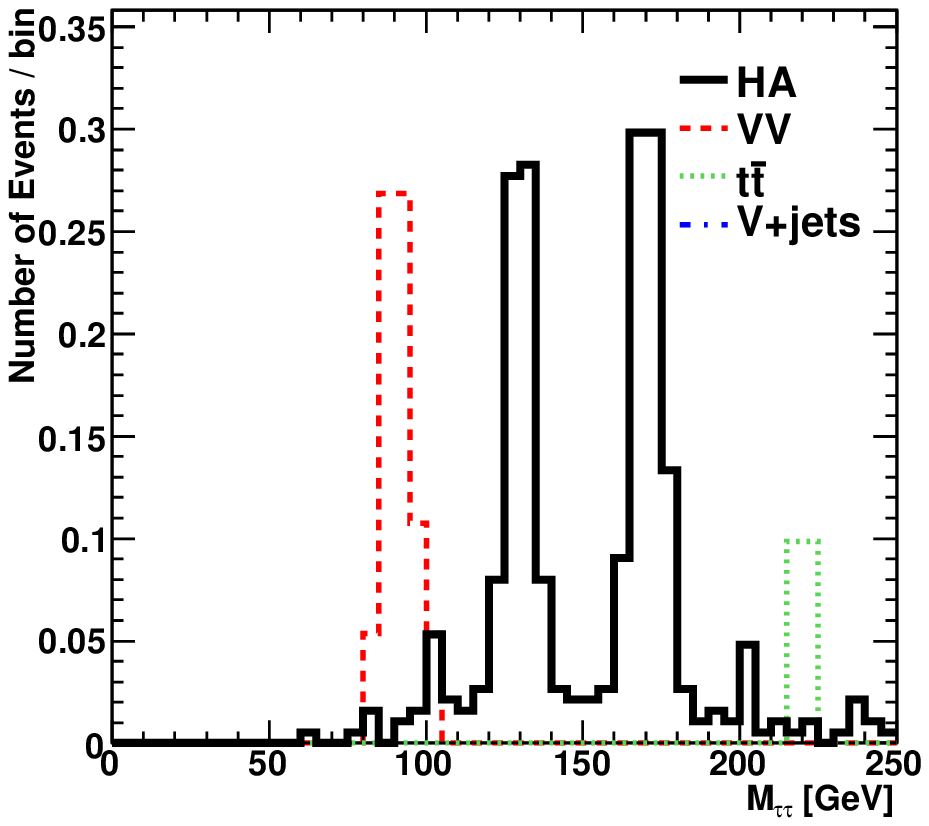}
 \caption{Invariant mass distributions of the tau-jet pair
 $M_{\tau_h\tau_h}$ (left) after the selection cuts up to the 
 opposite-charge muon pair in TABLE~\ref{Tab:LHC_2T2M}, and those of
 the tau-lepton pair reconstructed by using the collinear approximation
 $M_{\tau\tau}$ (right) in the $2\mu2\tau_h$ channel.
 Only the events with a solution of the collinear approximation in
 $0<z_{1,2}<1.1$ ranges are used for the latter distribution.
 Signal events plotted in the solid black line are generated with
 $m_H^{}=130$ GeV and $m_A^{}=170$ GeV.
 Contributions from the $VV$ background are also shown in the dashed red line.
 }
 \label{FIG:LHC_2T2M_Mtjtj}
\end{figure}

In the left panel in FIG.~\ref{FIG:LHC_2T2M_Mtjtj}, we show
the $M_{\tau_h\tau_h}$ distributions, after requiring opposite-sign for
the muon pair in TABLE~\ref{Tab:LHC_2T2M}.
In the right panel in FIG.~\ref{FIG:LHC_2T2M_Mtjtj}, the $M_{\tau\tau}$
distributions of the pair of the reconstructed tau leptons from the two 
tau-jets are shown, while the events without the solution of the
equations in the $0<z_{1,2}<1.1$ ranges are rejected. 
Similarly to the $M_{\mu\mu}$ distribution, two resonant peaks can
be observed with relatively wide widths at the masses of the neutral
Higgs bosons.
Only a few backgrounds from $VV$ and $t\bar t$ production are expected.
The $VV$ background can be further reduced by the cut on the
$m_Z^{}$-window for the reconstructed $M_{\tau\tau}$.

%
%
\begin{figure}[tb]
 \centering
 \includegraphics[height=7cm]{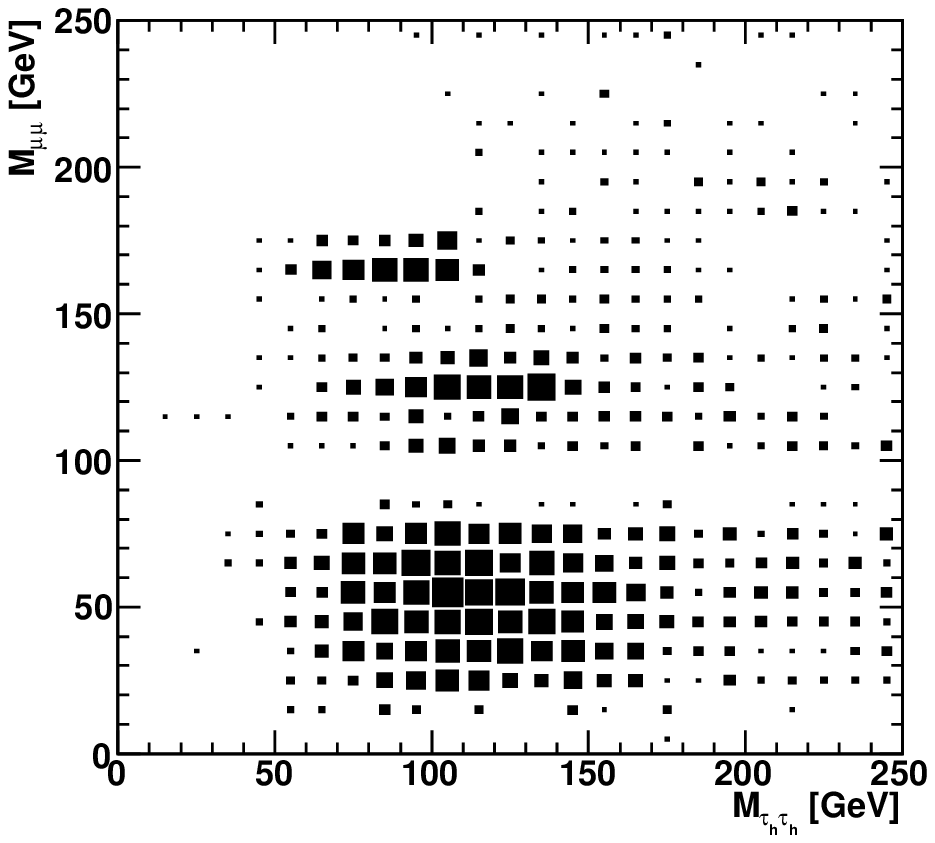}
 \includegraphics[height=7cm]{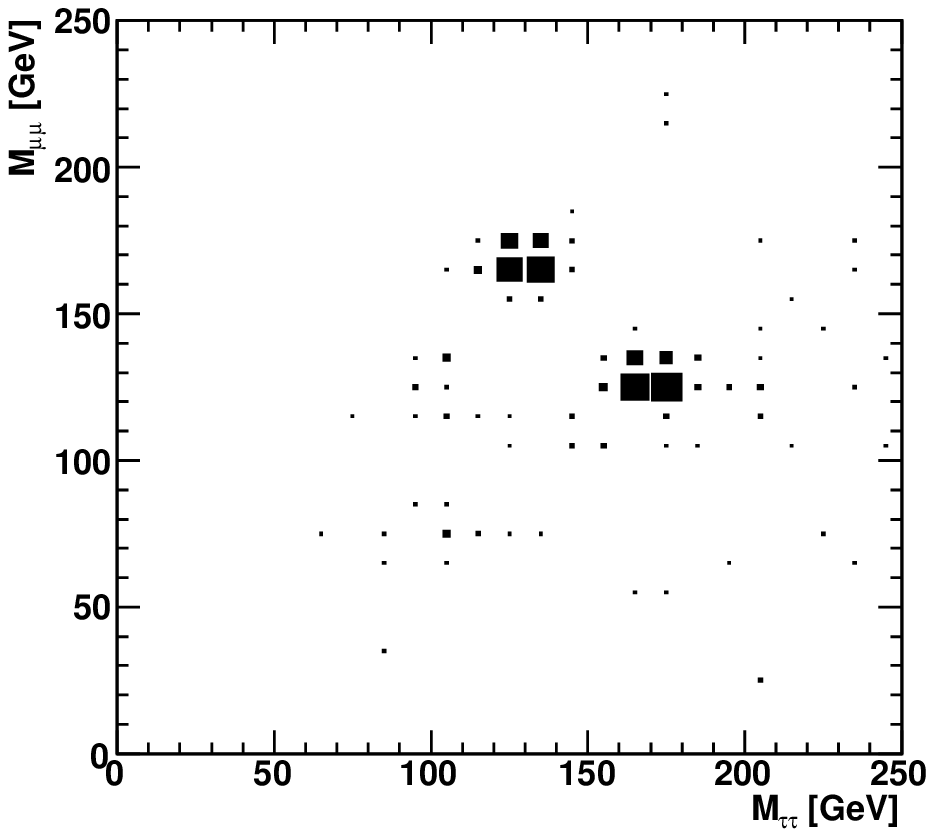} \\
 \includegraphics[height=7cm]{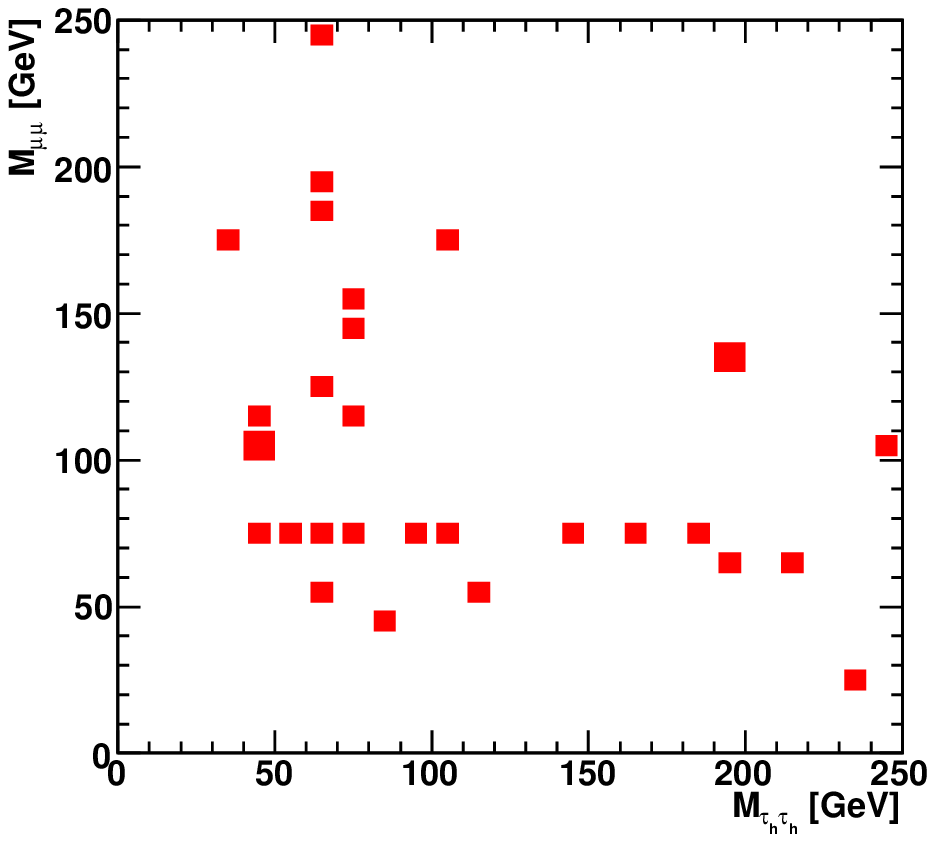}
 \includegraphics[height=7cm]{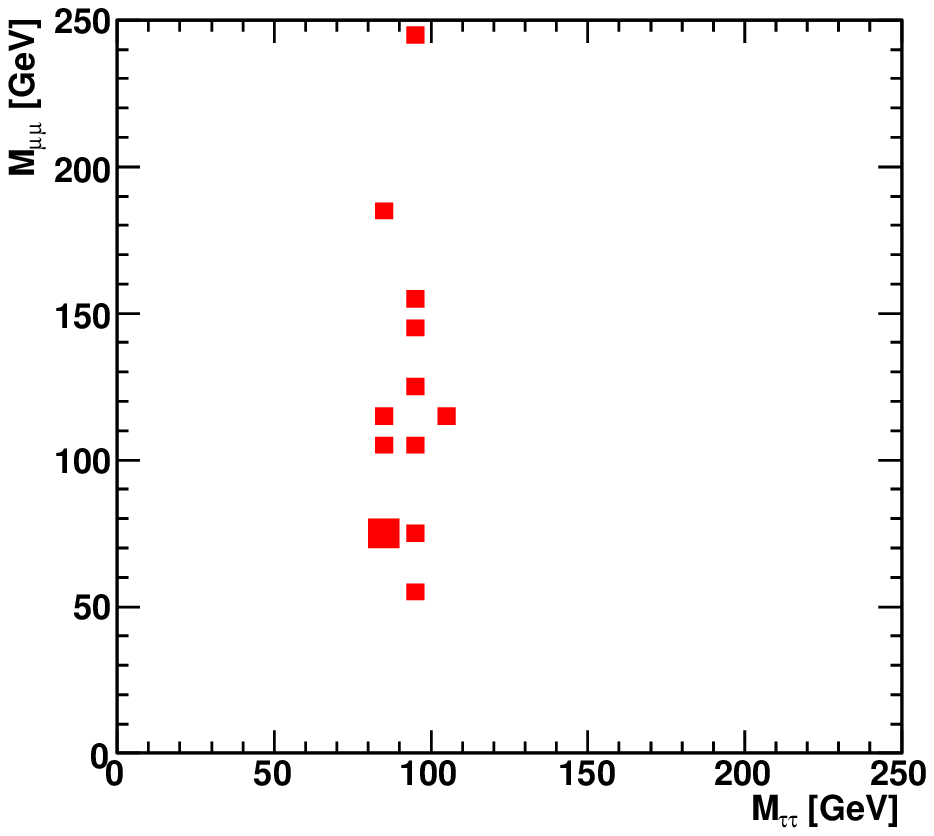}
 \caption{Two dimensional plots for the invariant mass distributions
 $M_{\mu\mu}$ vs. $M_{\tau_h\tau_h}$ (upper-left) and $M_{\mu\mu}$ vs.
 $M_{\tau\tau}$ (upper-right) in the signal processes $pp\to HA$, where
 $m_H^{}=130$ GeV and $m_A^{}=170$ GeV are taken.
 The distributions for the background events are also given in the lower panels.
 The normalization of the box-plot in each panel is set separately.
 }
 \label{FIG:LHC_2T2M_Mmumu_Mtjtj}
\end{figure}

In the upper panels of FIG.~\ref{FIG:LHC_2T2M_Mmumu_Mtjtj},
the double invariant mass distribution of
$M_{\tau_h\tau_h}\, \text{vs.}\, M_{\mu\mu}$ (left) and 
$M_{\tau\tau}\, \text{vs.}\, M_{\mu\mu}$ (right) for the signal events
are shown.
In the distribution in the top-left panel, we can see two bands along
the $x$-axis at $M_{\mu\mu}\simeq 130$~GeV and $170$~GeV. 
In the distribution in the top-right panel, two peaks can be found
around $(M_{\tau\tau},M_{\mu\mu})=(130, 170)$ and
$(170, 130)$ [GeV]. 
In the lower panels we show the distributions of the sum of the
background processes for the reference.
The background events after solving the tau lepton momenta mostly have
$M_{\tau\tau}\simeq m_Z$.
Thus by the $m_Z$-window cut on $M_{\tau\tau}$, more rejections of the
background are expected. 
This two dimensional invariant mass distribution gives an evidence for
the pair production of neutral Higgs bosons which dominantly decay into
the pair of tau leptons and occasionally into the pair of muons directly.
Accurate mass measurement is also possible, when we have sufficient
statistics. \\

\begin{table}[tb]
\centering
 \begin{tabular}{|c||cc|cc|cc|cc|cc|cc|}
  \hline
  \smash{\lower7pt\hbox{Lepton channels}}
  & \multicolumn{2}{c|}{$2\mu2\tau_h$}
  & \multicolumn{2}{c|}{$2\mu1\tau_h1e$}
  & \multicolumn{2}{c|}{$3\mu1\tau_h$}
  & \multicolumn{2}{c|}{$4\mu$}
  & \multicolumn{2}{c|}{$3\mu1e$}
  & \multicolumn{2}{c|}{$2\mu2e$} \\ [.3mm]
  & $s/b$ & ($S$) & $s/b$ & ($S$) & $s/b$ & ($S$) & $s/b$ & ($S$) &
  $s/b$ & ($S$) & $s/b$ & ($S$) \\
  \hline \hline
  Pre-selection
  & 98.7/16507. & (0.8) & 35.7/154. & (2.8) & 14.3/162. & (1.1) &
  0.9/401. & (0.0) & 2.6/8.9 & (0.8) & 3.3/647. & (0.1) \\
  $p_T^{\tau_h} > 40$~GeV
  & 42.8/618. & (1.7) & 23.6/46.2 & (3.2) & 9.4/46.8 & (1.3) & - & (-) &
  - & (-) & - & (-) \\
  $\cancel{E}_T^{} > 30$~GeV
  & 34.7/ 59.7 & (4.5) & 19.2/25.3 & (3.3) & 7.7/27.5 & (1.4) & 0.7/4.2
  & (0.3) & 2.0/5.8 & (0.8) & 2.5/5.7 & (1.0) \\
  $H_T^\text{jet} < 50$~GeV
  & 25.2/17.1 & (5.1) & - & (-) & - & (-) & - & (-) & - & (-) & - & (-)
  \\
  $H_T^\text{lep} > 250$~GeV
  & 22.7/13.3 & (5.1) & 15.7/18.4 & (3.3) & 6.5/20.2 & (1.4) & 0.6/2.1 &
  (0.4) & 1.5/2.8 & (0.8) & 1.7/3.0 & (0.9) \\
  $(m_Z^{})_{\mu\mu} \pm 10$~GeV
  & 19.8/6.4 & (5.8) & 13.7/2.3 & (5.9) & 5.4/5.1 & (2.1) & 0.5/0.5 &
  (0.6) & 1.3/0.6 & (1.3) & 1.3/0.3 & (1.6) \\
  \hline
  O.S.\ muons
  & 13.8/6.3 & (4.4) &  9.5/2.0 & (4.6) & - & (-) & - & (-) & - & (-) &
  0.9/0.3 & (1.3) \\
  $ 0 \le z_{1,2} \le 1.1$
  & 3.0/1.6 & (1.9) & 3.0/1.0 & (2.3) & 2.0/1.3 & (1.5) & 0.4/0.2 &
  (0.7) & 0.7/0.3 & (1.0) & 0.4/0.2 & (0.8) \\
  $(m_Z^{})_{\tau\tau} \pm 20$~GeV
  & 2.8/0.8 & (2.3) & 2.8/0.3 & (3.0) & 1.8/0.7 & (1.7) & 0.3/0.1 &
  (0.8) & 0.6/0.2 & (1.0) & 0.4/0.0 & (-) \\
  \hline
 \end{tabular}
 \caption{Summary of event rejections for the four-lepton channels with
 two or more muons. 
 The signal-to-background ratio $s/b$ and the significance $S$ for
 $L=100$~fb$^{-1}$ are listed. 
 The cut on $H_T^\text{jet}$ is not imposed for the channels with one or
 less tau-jet.
 A pairing rule of dimuons for the channels with three or more muons is
 explained in the text. 
 In the $2\mu2e$ channel, an additional cut of $|M_{ee}-m_Z^{}|>10$~GeV
 is applied at the cut of $(m_Z^{})_{\mu\mu}\pm 10$~GeV.
 }
 \label{Tab:LHC_2M}
\end{table}

Similar analysis to the $2\mu2\tau_h$ channel can be performed for
the other four-lepton channels with two or more muons, which are
$2\mu1\tau_h1e$, $3\mu1\tau_h$, $4\mu$, $3\mu1e$ and $2\mu2e$ channels.
In the channels where there are three or more muons, combinatorial
complexity may be avoided by choosing the pair of opposite-sign muons
which gives the largest transverse momentum of the muon pair.
The signal-to-background ratio and the expected significance for these
channels by applying the selection cuts similar to those for
the $2\mu2\tau_h$ channel are summarized in TABLE~\ref{Tab:LHC_2M} for
$L=100$~fb$^{-1}$.
The cut variables are not optimized for each channel.
The cut on $H_T^\text{jet}$ is not imposed for the channels with one or
less tau-jet, since for these channels $t\bar t$ process are negligible
from the pre-selection level.
The collinear approximation is applied for the two leptons other than the
two opposite-sign muons.
The large $s/b$ and $S$ can be observed for the $2\mu1\tau_h1e$
channel as well, where the $t\bar t$ background is less significant than
for the $2\mu2\tau_h$ channel.

By the requirement of opposite-sign charges in the dimuon, the
background processes hardly lose events, while the signal process
reduces one third of events which is consistent with a naive expectation
by counting combinatory. 
It means, in the other word, that the requirement of same-sign charges
in the dimuon extracts the signal events with only small background 
contribution mainly from the $VV$ process.
This is also useful to find the evidence on top of the SM process in the
various channels with two muons or also two electrons. 
Those channels also appear in the latter.

\subsubsection{Four-lepton channels with two or more tau-jets}

Due to the dominant branching ratio into tau leptons, a large number of
events is expected in the channels with high tau-jet multiplicity, thus
an excess beyond the SM expectation may be detectable in these channels.
On the other hand, four-lepton channels without a primary muon pair from
the decays of neutral Higgs bosons are hard to be kinematically
reconstructed. 

First, we consider the $4\tau_h$ channel. 
Since this channel contains only hadronic objects but no leptonic
objects in an event, the triggering efficiency may not be as good as those
in the channels with muons. 
However, the requirement of the high-$p_T^{}$ tau-jet shall stabilize
the abstraction of the events in this channel, and we expect at least
relative information would be available such as kinematical
distributions and the signal-to-background ratio.
Then, we present results and comments for the other four-lepton channels
with two or more tau-jets.
In the end of this subsection, we also comment on the remaining
four-lepton channels. \\

After the pre-selection with the requirement of the vanishing charge sum
of the four hadronic tau-jets, the expected numbers of events are 324,
147, 797 and 5105 for $HA$, $VV$, $t\bar t$ and $V+$jets production
processes, respectively, for $L=100$~fb$^{-1}$. 
The dominant background contributions come from $Z+$jets production
followed by the $Z\to\tau^+\tau^-$ decay and the hadronic decays of the
tau leptons with two more tau-jets which are mis-identified from the ISR
jets.

\begin{figure}[tb]
 \includegraphics[height=5.3cm]{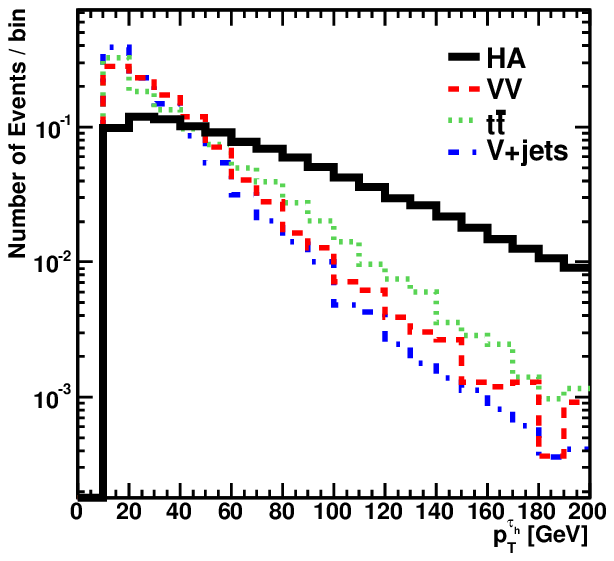}
 \includegraphics[height=5.3cm]{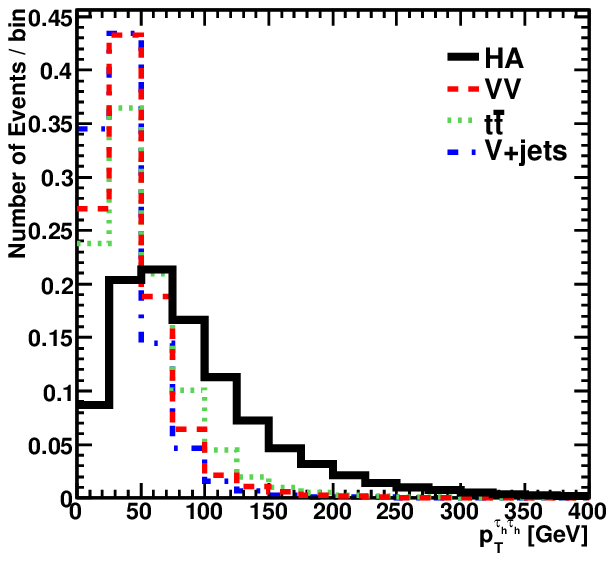} \hspace{5.3cm}\,
 \includegraphics[height=5.3cm]{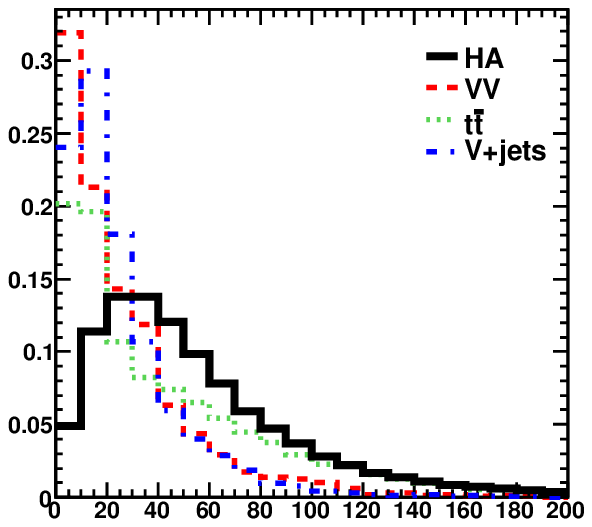}
 \includegraphics[height=5.3cm]{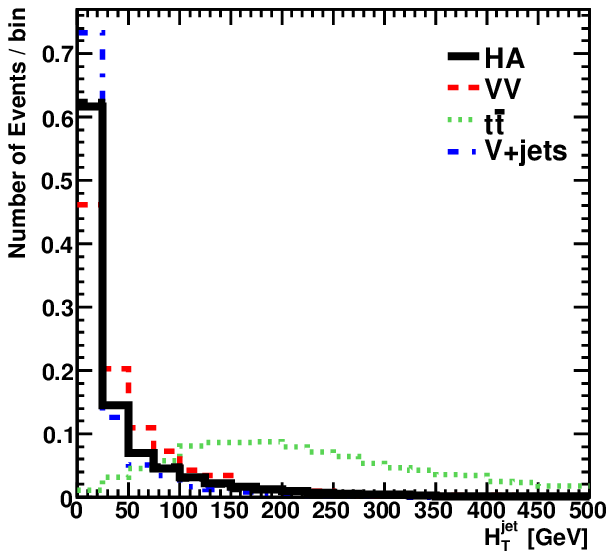}
 \includegraphics[height=5.3cm]{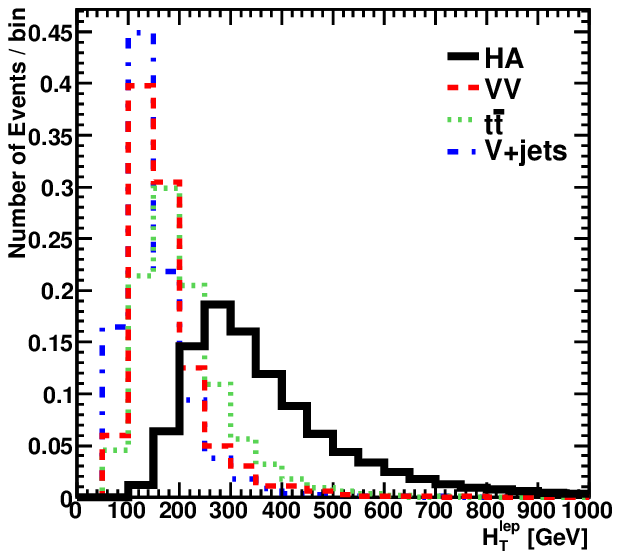}
\caption{Distributions of kinematical variables for the $4\tau_h$ channel.
 Line descriptions follow those in FIG.~\ref{FIG:LHC_2T2M_Mmumu}.}
\label{FIG:Dist_LHC_4T}
\end{figure}
\begin{table}[tb]
\begin{center}
\begin{tabular}{|c||c|c||c|c|c||c|c|}
 \hline $4\tau_h$ event analysis & $HA$ & $\phi^0 H^{\pm}$ & $VV$ &
 $t\bar t$ & $V+$jets & $s/b$ & $S$ (100~fb$^{-1}$) \\ \hline \hline
 Pre-selection
 & 324. &  52.8 & 147.  & 797. &  5105.  & 0.1 &   4.7 \\
 $p_T^{\tau_h} > 40$~GeV
 &  67.2 &  4.9 &   2.0 &  14.7 &   21.7 & 1.9 &   9.4 \\
 $\cancel{E}_T^{} > 30$~GeV
 &  48.6 &  4.4 &   1.1 &   7.6 &   10.4 & 2.8 &   9.3 \\
 $H_T^\text{jet} < 50$~GeV
 &  34.2 &  3.4 &   0.5 &   0.8 &    8.2 & 3.9 &   8.7 \\
 $H_T^\text{lep} > 350$~GeV
 &  27.6 &  2.7 &   0.4 &   0.5 &    3.1 & 7.5 &   9.3 \\
 \hline
\end{tabular}
\caption{Table for the background reductions in the $4\tau_h$ channel.}
\label{Tab:LHC_4T}
\end{center}
\end{table}

In FIG.~\ref{FIG:Dist_LHC_4T}, we show distributions of the transverse
momentum $p_T^{\tau_h}$ of the tau-jet, that $p_T^{\tau_h\tau_h}$ for
the tau-jet pair, the missing transverse energy $\cancel{E}_T$ and 
the scalar sum of the hadronic and leptonic transverse momentum,
$H_T^\text{jet}$ and $H_T^\text{lep}$, respectively.
Distributions for all the signal and background processes are normalized 
to be unity. 
Line descriptions follow those in FIG.~\ref{FIG:LHC_2T2M_Mmumu}.
In order to reject backgrounds, the following selection cuts are applied;
\begin{subequations}
\begin{align}
 & p_{T}^{\tau_h}        >  40~\text{GeV}, \\
 & \cancel{E}_T^{}       >  30~\text{GeV}, \\
 & H_T^\text{jet}        <  50~\text{GeV}, \\
 & H_T^\text{lep}        > 350~\text{GeV}.
\end{align}
\end{subequations}
Reductions of the signal and background events are summarized at each cut
in TABLE~\ref{Tab:LHC_4T}.
For $L=100$~fb$^{-1}$, an expected significance $S$ is about $5$ at the
pre-selection level, and reaches to around $9$ after the selection
cuts. 
The signal-to-background ratio $s/b$ rises to about 8 at the end of all
the selection cuts.
Thus, finding the evidence in this channel seems very promising.
It may be even possible at more earlier stage at the LHC. 
The more tight cut for the $p_T^{}$ of the tau-jets can enhance the 
signal-to-background ratio, while slightly reduces the significance.
We note that we have not estimated the background contributions from pure
QCD processes as well as the $V+$jet production process followed by
hadronic decays of weak gauge bosons, which could survive after the
selection cuts if all the four tau-jets are mis-identified.
We expect that a severe $\cancel{E}_T$ cut should sufficiently reduce the
number of events for these processes. \\

Since in the $4\tau_h$ events all the four hadronic decays of tau leptons 
could yield missing momenta, the collinear approximation analysis used 
in the $2\mu2\tau_h$ channel cannot be applied, and the masses of the
Higgs bosons cannot be reconstructed kinematically.
However, as we will see below, distributions of the invariant mass of
the tau-jet pair are useful to obtain the information of the mass of the
Higgs bosons.
To resolve combinatorial complexity from the four tau-jets, one
invariant mass is constructed from the tau-jet pair with
opposite-charges which gives the largest $p_T^{\tau_h\tau_h}$, and the
other is constructed from the remaining tau-jet pair.
Then, the invariant masses of the two tau-jet pairs, $M_{\tau_h\tau_h}$
and $m_{\tau_h\tau_h}$, are defined by $M_{\tau_h\tau_h} \ge
m_{\tau_h\tau_h}$.

\begin{figure}[tb]
 \centering
 \includegraphics[height=7cm]{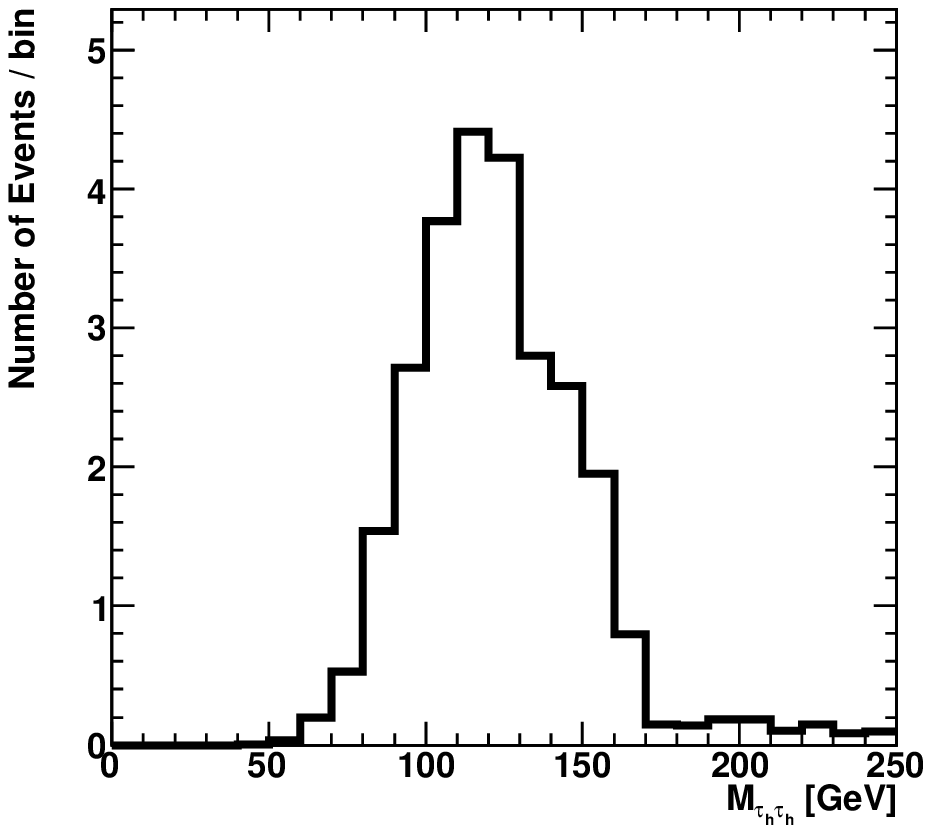}
 \includegraphics[height=7cm]{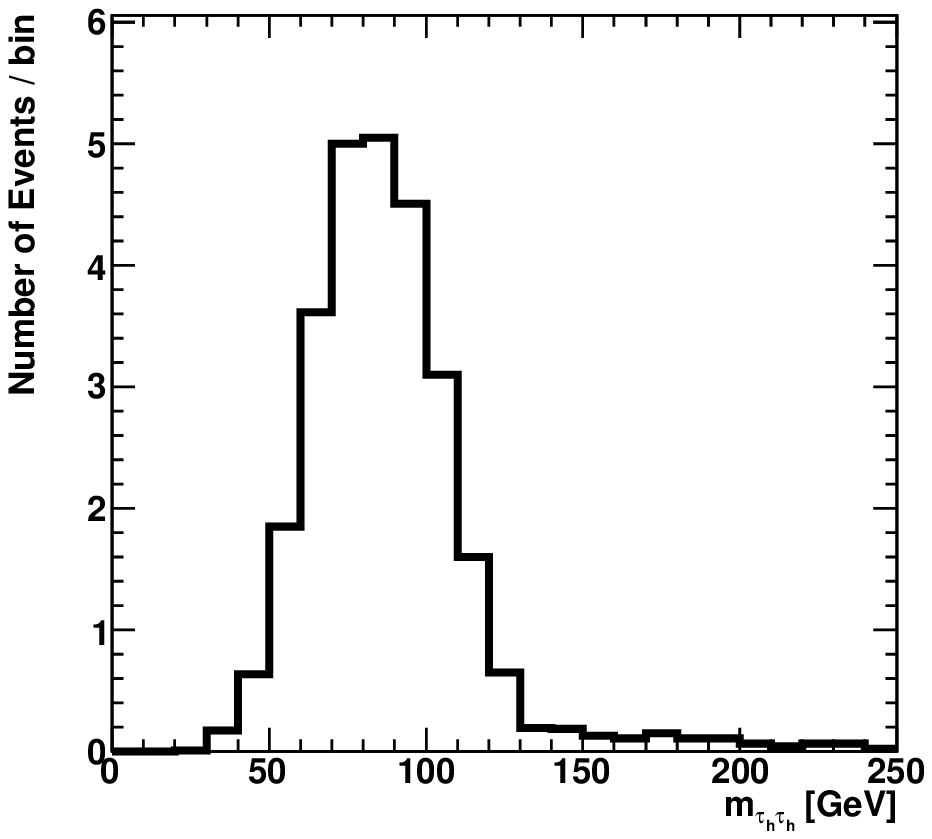}
 \caption{Distributions of the invariant mass of the opposite-charge
 tau-jet pair, $M_{\tau_h\tau_h}$ (left) and $m_{\tau_h\tau_h}$ (right),
 where $M_{\tau_h\tau_h}\ge m_{\tau_h\tau_h}$ in the $4\tau_h$
 channel.
 Signal events are generated with $m_H^{}=130$~GeV and $m_A^{}=170$~GeV.}
 \label{FIG:LHC_4T_Mtjtj}
\end{figure}
\begin{figure}[tb]
 \centering
 \includegraphics[height=7cm]{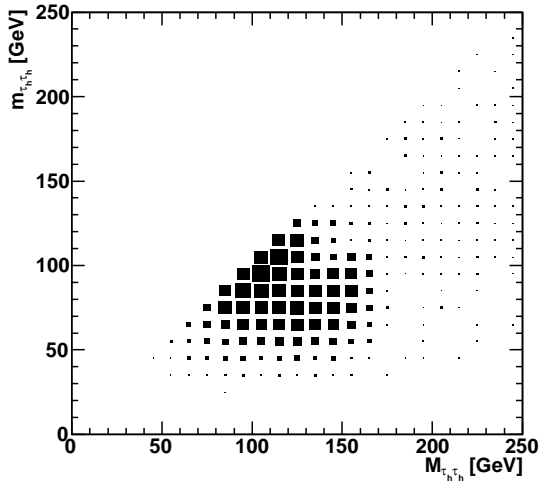}
 \caption{Two dimensional distributions of the two opposite-sign tau-jet
 pair ($M_{\tau_h\tau_h}$ vs $m_{\tau_h\tau_h}$) in the $4\tau_h$
 channel after the selection cuts for the signal processes $pp\to HA$
 with $m_H^{}=130$~GeV and $m_A^{}=170$~GeV.}
 \label{FIG:LHC_4T_Mtjtj1_Mtjtj2}
\end{figure}

In FIG.~\ref{FIG:LHC_4T_Mtjtj}, distributions of $M_{\tau_h\tau_h}$ (left) 
and $m_{\tau_h\tau_h}$ (right) are shown for the signal events from $HA$
production, while the background contributions are omitted because they
are negligible.
In the $M_{\tau_h\tau_h}$ distribution in the left panel, we can see an
endpoint of the distribution at $M_{\tau_h\tau_h}\simeq 170$~GeV which
corresponds to $m_A^{}=170$~GeV.
Another mass information for $m_H^{}=130$~GeV can be seen as an endpoint
in the $m_{\tau_h\tau_h}$ distribution in the right panel or may be
found as a bump in the $M_{\tau_h\tau_h}$ distribution.
Thus, if sufficient numbers of events are obtained by accumulating the
integrated luminosity such like a few thousand~fb$^{-1}$ at the LHC, it
should be possible to extract the mass of the Higgs bosons by these
distributions. 
In FIG.~\ref{FIG:LHC_4T_Mtjtj1_Mtjtj2}, we also present the two
dimensional distribution of the invariant masses $M_{\tau_h\tau_h}$ and
$m_{\tau_h\tau_h}$ for the signal events in the $4\tau_h$ channel after
the selection cuts. \\

Similar analysis to the $4\tau_h$ channel can be performed for the
other four-lepton channels with two or more hadronic tau-jets; such as
$3\tau_h1\mu$, $3\tau_h1e$, $2\tau_h1\mu1e$ and $2\tau_h2e$ channels.
Due to the expected large number of events, many of them can give large
$S$ with large $s/b$ by the similar selection cuts for
$L=100$~fb$^{-1}$. 
Measurements of these channels would provide a test of the signatures
of tau leptons in the extensive channels, and thus they are useful
to probe the production of the tau-lepton-specific Higgs bosons.
The endpoint study such like in FIG.~\ref{FIG:LHC_4T_Mtjtj} may not be
suitable for the invariant mass distributions including $\mu$ or $e$,
because the momentum fraction $z$ for the leptonic decays of tau leptons
distributes mainly in a small $z$ region~\cite{Ref:BHM}. 
The rest of the four-lepton channels are $1\tau_h1\mu2e$, $1\tau_h3e$,
$1\mu3e$ and $4e$.
However, except the $1\tau_h1\mu2e$ channel, the expected numbers of
events are too small; a few events or less for $L=100$~fb$^{-1}$.
The signal-to-background ratio and the estimated significance by
applying the selection cuts for these channels are summarized in
TABLE~\ref{Tab:LHC_2T} and \ref{Tab:LHC_1T} assuming the integrated
luminosity of 100~fb$^{-1}$. \\

Before closing the subsection for the four-lepton channels, we point out
an interesting comparison between the $2\mu2\tau_h$ and $2\tau_h2e$
channels. 
Theoretically, the difference of the two channels arises from the direct
decay of Higgs bosons into dimuons through the Yukawa couplings of Higgs
bosons to muons.
On the other hand, those to electrons are negligible, since
$m_e^2/m_{\mu}^2\simeq 2.3\cdot 10^{-5}$. 
The invariant mass distribution of the dielectron in the $2\tau_h2e$ 
channels is nothing but that of the dimuon in the $2\tau_h2\mu$ channel
except the absence of the two resonant peaks seen at $M_{\mu\mu}\simeq
m_{H}^{}$ and $M_{\mu\mu}\simeq m_{A}^{}$.
The common part of the distributions between the $2\mu2\tau_h$ and
$2\tau_h2e$ channels should be originated from the $4\tau$ decay, while
the excess in the $2\mu2\tau_h$ channel relative to the $2\tau_h2e$
channel should be originated from the decay of the Higgs bosons directly
into muons.

Experimentally, the efficiencies of finding electrons and muons are
different due to the different isolation condition and acceptance cuts.
In our simulation study, the normalization of the events could be
corrected by naively multiplying for the number of events in
the $2\tau_h2e$ channel by $1.1$, which is about the ratio of the
finding efficiency for muons to that for electrons obtained by comparing
the numbers of events in the $3\tau_h1\mu$ and $3\tau_h1e$ channels. 

\begin{table}[tb]
  \centering
  \begin{tabular}{|c||cc|cc|cc|cc|cc|}
  \hline
  \smash{\lower7pt\hbox{Lepton channels}}
  & \multicolumn{2}{c|}{$4\tau_h$}
  & \multicolumn{2}{c|}{$3\tau_h1\mu$}
  & \multicolumn{2}{c|}{$3\tau_h1e$}
  & \multicolumn{2}{c|}{$2\tau_h1\mu1e$}
  & \multicolumn{2}{c|}{$2\tau_h2e$}
  \\ [.3mm]
   & $s/b$ & ($S$) & $s/b$ & ($S$) & $s/b$ & ($S$) & $s/b$ & ($S$) &
   $s/b$ & ($S$) \\
   \hline \hline
   Pre-selection
   & 377./6050. & (4.8) & 302./4208. & (4.6) & 278./3883. & (4.4) &
   166./917. & (5.3) & 74.4/13202. & (0.6) \\
   $p_T^{\tau_h} > 40$~GeV
   & 72.1/38.5 & (9.5) & 87.2/70.2 & (8.9) & 80.2/72.2 & (8.2) &
   71.7/67.5 & (7.6) & 32.4/479. & (1.5) \\
   $\cancel{E}_T^{} > 30$~GeV
   & 53.0/19.0 & (9.3) & 69.3/54.6 & (8.0) & 63.4/53.8 & (7.5) &
   58.0/58.6 & (6.7) & 26.3/38.6 & (3.8) \\
   $H_T^\text{jet} < 50$~GeV
   & 37.6/9.6 & (8.7) & 49.0/17.4 & (8.9) & 44.9/23.0 & (7.6) &
   41.7/13.7 & (8.5) & 18.7/16.0 & (4.0) \\
   $H_T^\text{lep} > 350$~GeV
   & 30.3/4.0 & (9.3) & 34.5/8.4 & (8.4) & 31.4/10.9 & (7.2) &
   24.2/3.8 & (8.0) & 10.7/8.2 & (3.2) \\
   $(m_Z)_{ee}\pm 10$~GeV
   & - & (-) & - & (-) & - & (-) & - & (-) & 9.3/2.5 & (4.2) \\
  \hline
 \end{tabular}
 \caption{Summary of event rejections by the selection cuts for the
  four-lepton channels with two or more tau-jets.}
  \label{Tab:LHC_2T}
  \begin{tabular}{|c||cc|cc|cc|cc|}
  \hline
  \smash{\lower7pt\hbox{Lepton channels}}
  & \multicolumn{2}{c|}{$1\tau_h1\mu2e$}
  & \multicolumn{2}{c|}{$1\tau_h3e$}
  & \multicolumn{2}{c|}{$1\mu3e$}
  & \multicolumn{2}{c|}{$4e$}
  \\[.3mm]
  & $s/b$ & ($S$) & $s/b$ & ($S$) & $s/b$ & ($S$) & $s/b$ & ($S$) \\
  \hline \hline
  Pre-selection
  & 29.2/132. & (2.5) & 8.7/120. & (0.8) & 1.7/7.6 & (0.6) & 0.4/268. &
   (0.0) \\
  $p_T^{\tau_h} > 40$~GeV
  & 19.3/38.6 & (2.9) & 5.6/34.2 & (0.9) & - & (-) & - & (-) \\
  $\cancel{E}_T^{} > 30$~GeV
  & 15.5/22.1 & (3.0) & 4.6/19.2 & (1.0) & 1.2/3.4 & (0.6) & 0.3/2.6 &
   (0.2) \\
  $(m_Z)_{ee}\pm 10$~GeV
  & 13.6/2.4 & (5.8) & 4.0/6.5 & (1.4) & 1.1/1.2 & (0.9) & 0.2/0.7 &
   (0.2) \\
  \hline
  \end{tabular}
  \caption{Summary of event rejections by the selection cuts for the rest of
  the four-lepton channels.}
  \label{Tab:LHC_1T}
\end{table}
%

\subsection{Charged Higgs boson associated production}

At the hadron collider, charged Higgs bosons can be produced in
association with the neutral Higgs bosons via the $pp \to W^{\pm\,*} \to
\phi^0 H^\pm$ processes~\cite{AH+}.
In the Type-X THDM, more than $99\%$ of charged Higgs bosons decays into
a tau lepton and a neutrino, and $0.35\% \simeq m_\mu^2/m_\tau^2$ does
into a muon and a neutrino for $\tan\beta\gtrsim 2$~\cite{Ref:AKTY}.
Therefore, the characteristic signatures for this process would be
the three-lepton channels including many tau leptons or muons
accompanied with large missing momentum. 

There are ten kinds of the three-lepton channels.
Following the analysis of the four-lepton channels, we first present an
analysis for the three-lepton channels with two or more muons, then we
present analysis for those with two or more tau-jets.
Brief summary of the analysis for the rest of the three-lepton channels
is presented after a while.

\subsubsection{Three-lepton channels with two or more muons}
\begin{figure}[tb]
 \centering
 \includegraphics[height=7cm]{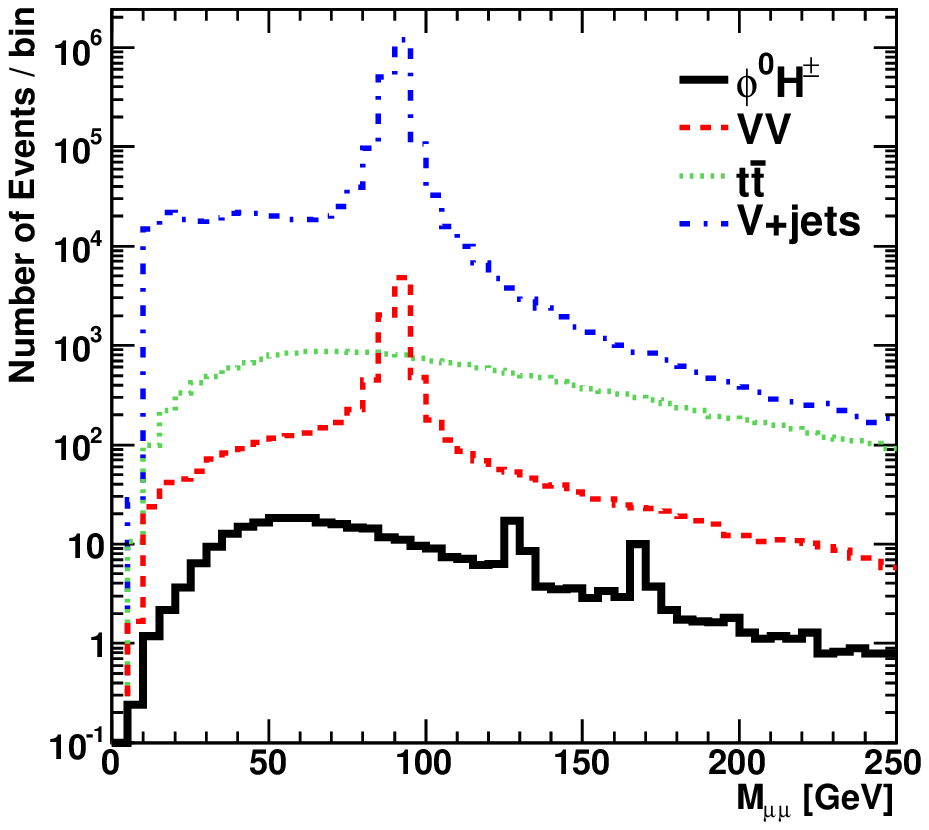}
 \includegraphics[height=7cm]{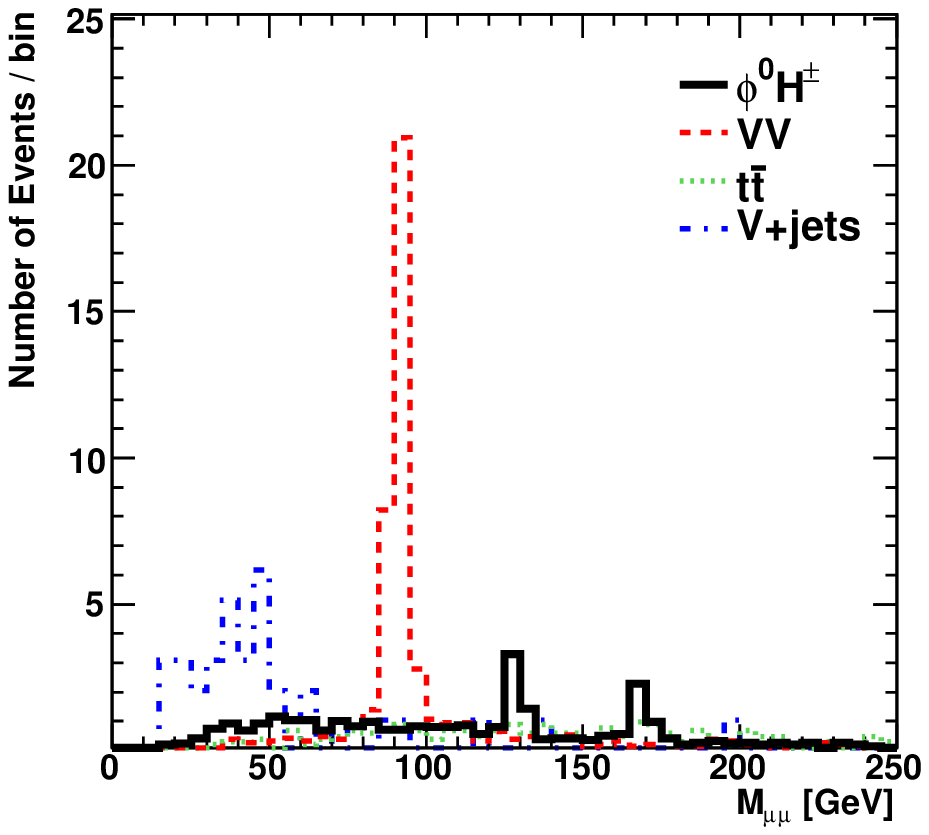}
 \caption{The dimuon invariant mass distribution for the $2\mu1\tau_h$
 channel after the pre-selection (left) and after the selection cuts
 before the cut on the $m_Z^{}$-window in TABLE~\ref{Tab:LHC_1T2M} (right).
 For the signal process, $m_H^{}=130$~GeV, $m_A^{}=170$~GeV and
 $m_{H^\pm}=150$~GeV are taken.
 }
 \label{FIG:LHC_1T2M_Mmumu}
\end{figure}

In the $2\mu1\tau_h$ channel, the two muons can be originated from the
decay of $H$ or $A$, or one muon comes directly from $H^\pm$ and the
other comes secondarily through the leptonic decay of the tau leptons.
In the former case, the two muons must have opposite charges, and the
distribution of their invariant mass shows resonance peaks at the mass
of neutral Higgs bosons.

In the left panel in FIG.~\ref{FIG:LHC_1T2M_Mmumu}, we plot the
invariant mass distribution of the muon pair for the $2\mu1\tau_h$ channel
after the pre-selection.
The distributions are scaled to the expected number of the events for
$L=100$~fb$^{-1}$ for each signal and background process. 
The expected number of events from the signal $\phi^0 H^{\pm}$ (and
$HA$) process is four orders of magnitude smaller than that for the
background processes, where the dominant contributions come from the
$Z+$jets process. 

\begin{figure}[tb]
 \includegraphics[height=5.3cm]{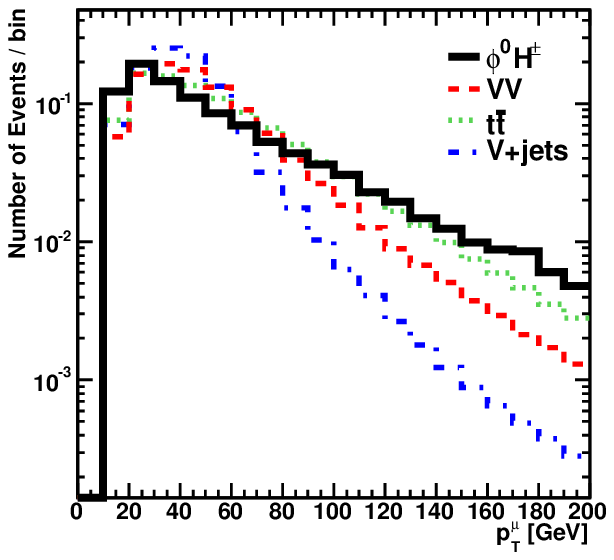}
 \includegraphics[height=5.3cm]{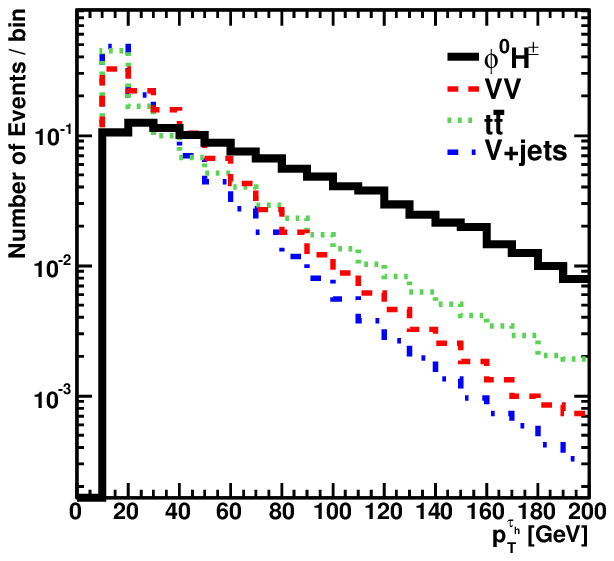}
 \includegraphics[height=5.3cm]{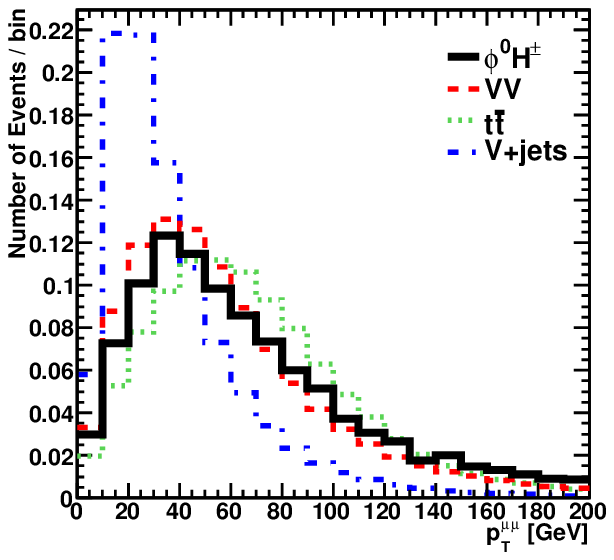}
 \includegraphics[height=5.3cm]{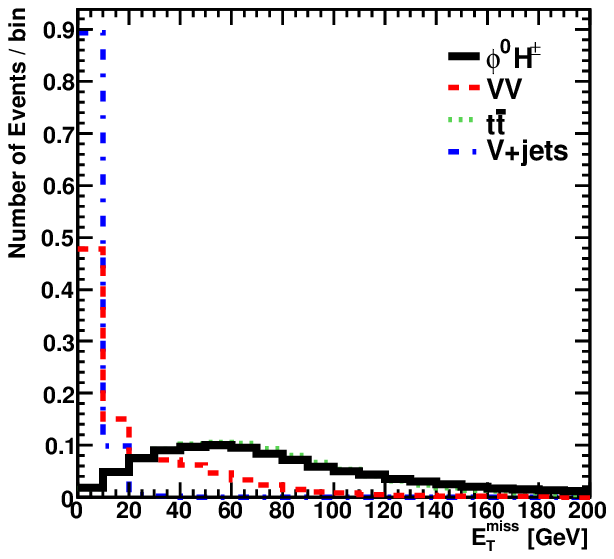}
 \includegraphics[height=5.3cm]{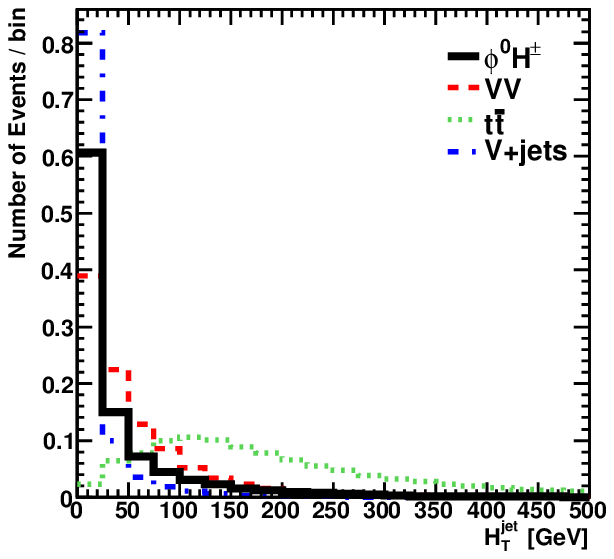}
 \includegraphics[height=5.3cm]{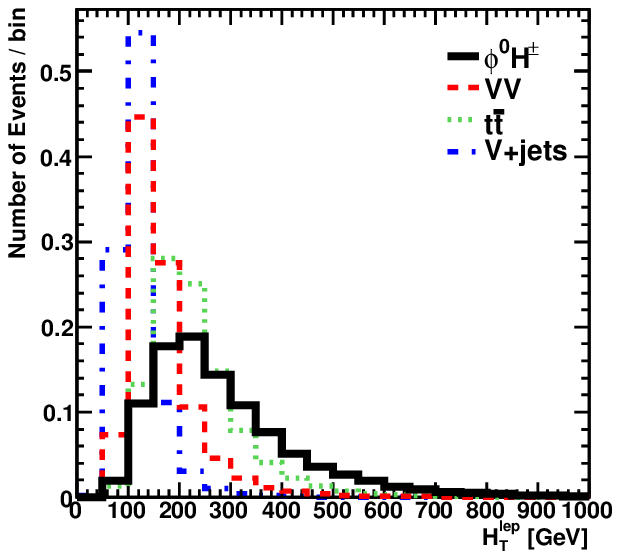}
 \caption{Distributions of kinematical variables for the $2\mu1\tau_h$ channel.
 Line descriptions follow those in FIG.~\ref{FIG:LHC_2T2M_Mmumu}.}
 \label{FIG:Dist_LHC_1T2M}
\end{figure}

To extract the signal events, we consider the selection cuts on this
channel.
In FIG.~\ref{FIG:Dist_LHC_1T2M}, we plot distributions of the transverse
momentum $p_T^\mu$ for a muon, that $p_T^{\tau_h}$ for a tau-jet and that 
$p_T^{\tau_h\tau_h}$ for a tau-jet pair, the missing transverse energy
$\cancel{E}_T$, and the scalar sum $H_T^\text{jet}$ and $H_T^\text{lep}$ 
of hadronic and leptonic transverse momentum, respectively,
in the $2\mu1\tau_h$ events after the pre-selection.
Looking at the distributions in FIG.~\ref{FIG:Dist_LHC_1T2M}, 
we perform the following selection cuts to extract signal events 
in the $2\mu1\tau_h$ channels;
\begin{subequations}
\begin{align}
 & p_T^{\tau_h}     >  40~\text{GeV}, \\
 & \cancel{E}_T^{}  >  80~\text{GeV}, \\
 & H_T^\text{jet}   <  30~\text{GeV}, \\
 & H_T^\text{lep}   > 350~\text{GeV}, \\
 & |M_{\mu\mu}-m_Z| >  10~\text{GeV}.
\end{align}
\end{subequations}
Reductions of the number of events are summarized in each step in
TABLE~\ref{Tab:LHC_1T2M}.
\begin{table}[tb]
\begin{center}
\begin{tabular}{|c||c|c||c|c|c||c|c|}
 \hline $2\mu1\tau_h$ event analysis & $HA$ & $\phi^0 H^{\pm}$ & $VV$ &
 $t\bar t$ & $V+$jets & $s/b$ & $S$ (100~fb$^{-1}$) \\ \hline \hline
 Pre-selection
 & 103.  & 341.  & 10380.  & 22683.  & 2297490.  & $10^{-4}$ &  0.3 \\
 $p_T^{\tau_h} > 40$~GeV
 &  71.4 & 230.  &  3147.  &  6690.  &  455178.  & $10^{-3}$ &  0.4 \\
 $\cancel{E}_T^{} > 80$~GeV
 &  25.0 & 110.  &   244.  &  2659.  &     260.  & $10^{-2}$ &  2.4 \\
 $H_T^\text{jet} < 30$~GeV
 &  10.3 &  62.8 &   122.  &    96.5 &     102.  & 0.2 &  3.9 \\
 $H_T^\text{lep} > 350$~GeV
 &   5.9 &  33.8 &    50.8 &    26.0 &      36.0 & 0.4 &  3.5 \\
 $(m_Z^{})_{\mu\mu} \pm 10$~GeV
 &   5.4 &  30.8 &    17.4 &    23.0 &      34.0 & 0.5 &  3.9 \\
 \hline
 O.S.\ muons
 &   4.1 &  21.3 &    16.3 &    23.0 &      34.0 & 0.3 &  2.8 \\
 $p_T^{\mu\mu} > 80$~GeV
 &   2.7 &  14.4 &    10.5 &    12.4 &      18.5 & 0.4 &  2.5 \\
 $(m_{\phi^{0}}^{})_{\mu\mu} \pm 10$~GeV
 &   0.7 &   5.4 &     1.9 &     2.7 &       1.0 & 1.1 &  2.3 \\
 \hline
\end{tabular}
\caption{Table for background reductions in the $2\mu1\tau_h$ channel.}
\label{Tab:LHC_1T2M}
\end{center}
\end{table}
After the $m_Z^{}$-window cut on the $M_{\mu\mu}$ distribution, the
number of events for signal processes is $\sim 36$, while the number of
background events is expected to be $\sim 74$. 
Although the $s/b$ ratio is less than one, the significance $S$ reaches
to $\sim 4$ for $L=100$~fb$^{-1}$.
Therefore, if a reliable background estimation is achieved, the excess
of the observed events could be the signature for the production of the
neutral and charged Higgs bosons. 

In the right panel in FIG.~\ref{FIG:LHC_1T2M_Mmumu}, we plot the
$M_{\mu\mu}$ distribution after the selection cuts up to the cut on
$H_T^\text{lep}$. 
After the selection cuts, the $M_{\mu\mu}$ distribution in the $V+$jets
background events tends to drop quickly before $M_{\mu\mu}\simeq50$~GeV. 
This is due to the requirement of the large $\cancel{E}_T$, thus the events
with a muon pair which directly comes from the decay of $Z$ bosons are
mostly rejected, but the events with a muon pair from the $Z\to\tau^+\tau^-$
decay followed by the muonic decays of the tau leptons remain with one
mis-identified tau-jet in the $2\mu1\tau_h$ channel. 
The $VV$ backgrounds mostly come from $WZ$ production followed by
$Z\to\mu^+\mu^-$ and $W^{\pm}\to\tau^{\pm}\nu$ decays.
Thus, the $M_{\mu\mu}$ distribution has a peak at $M_{\mu\mu}\simeq
m_Z^{}$, and the $VV$ background events are reduced by the
$m_Z^{}$-window cut significantly. 
The signal events from $\phi^0 H^{\pm}$ production could show up sharp
peaks at $M_{\mu\mu}\simeq m_H^{}$ and $M_{\mu\mu}\simeq m_A^{}$ on top
of the background contributions.
The resolution would be sufficient to determine the mass of the Higgs
bosons, since it is constructed from the muon momenta.

In order to measure the mass of the charged Higgs boson, we propose an
analysis using the transverse mass variable. 
First, we further purify the events by requiring opposite-sign
in the muon pair to extract the dimuon which comes from the decay of 
the neutral Higgs boson.
For the same purpose, the cut on the transverse momentum of the dimuon is
applied.
This cut reduces the significance $S$ slightly, however, it rather
enhances the signal-to-background ratio $s/b$. 
Such purification of the signal events enables us to see the delicate
structure of the transverse mass distribution for the signal events.
The transverse mass of the tau-jet and the missing transverse momentum
is constructed by
\begin{align}
 M_T = \sqrt{2\,p_T^{\tau_h}\cdot\cancel{E}_T\cdot(1-\cos\Delta\phi)},
\label{Eq:MT}
\end{align}
where $\Delta\phi$ is the difference of azimuthal angles of the tau-jet
and the missing transverse momentum.

\begin{figure}[tb]
 \centering
 \includegraphics[height=7cm]{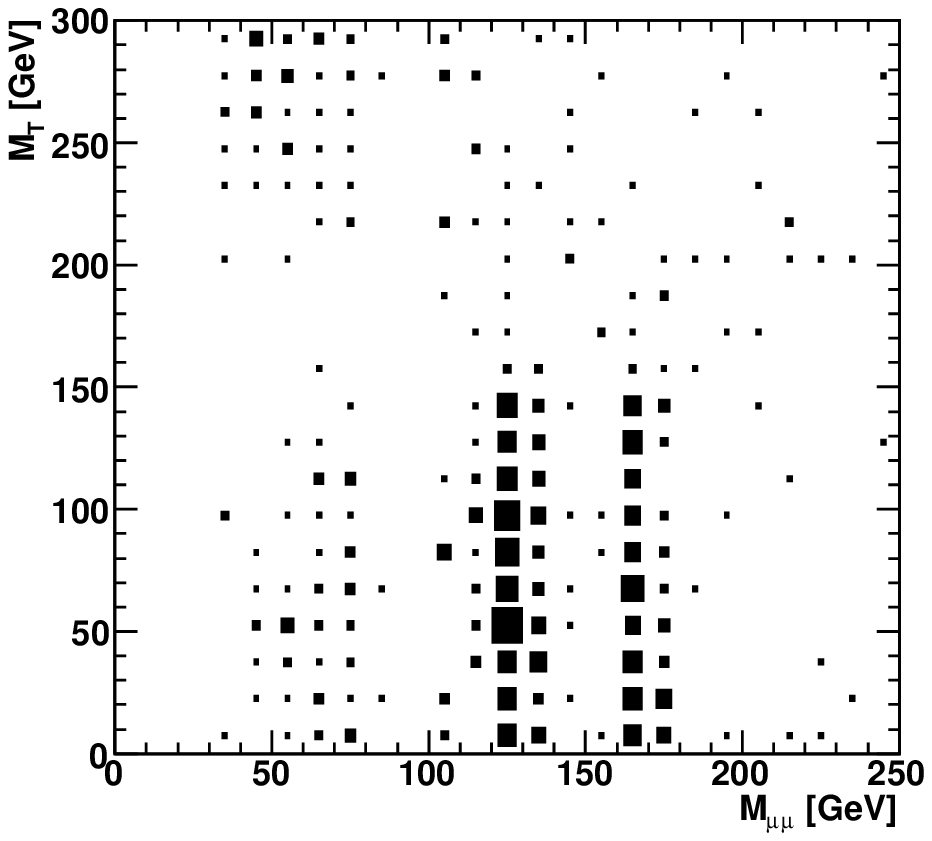}
 \includegraphics[height=7cm]{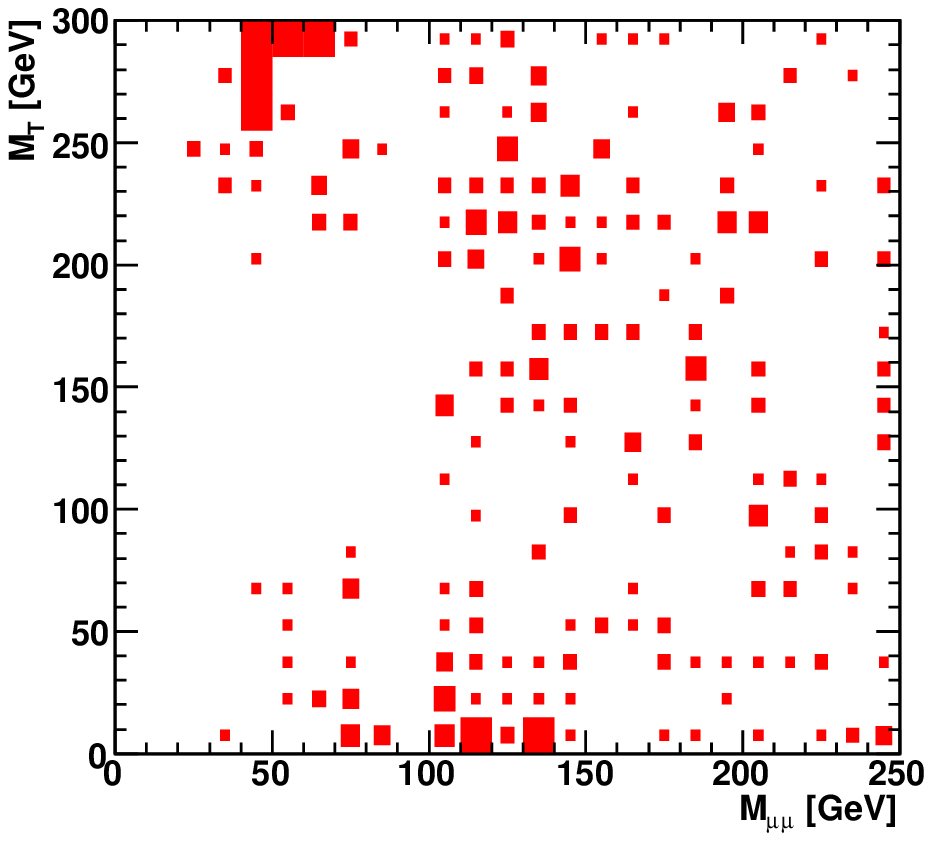}
 \caption{Two dimensional distributions of the dimuon invariant mass and
 the transverse mass of the tau-jet and the missing transverse momentum
 for the signal process (left) and the sum of the background processes
 (right) in the $2\mu1\tau_h$ channel after the all selection cuts up to
 $p_T^{\mu\mu}>80$~GeV in TABLE~\ref{Tab:LHC_1T2M}.
The signal process is simulated with $m_H^{}=130$ GeV, $m_A^{}=170$ GeV and
 $m_{H^\pm}^{}=150$ GeV.}
 \label{FIG:MT}
\end{figure}

\begin{figure}[tb]
 \centering
 \includegraphics[height=7cm]{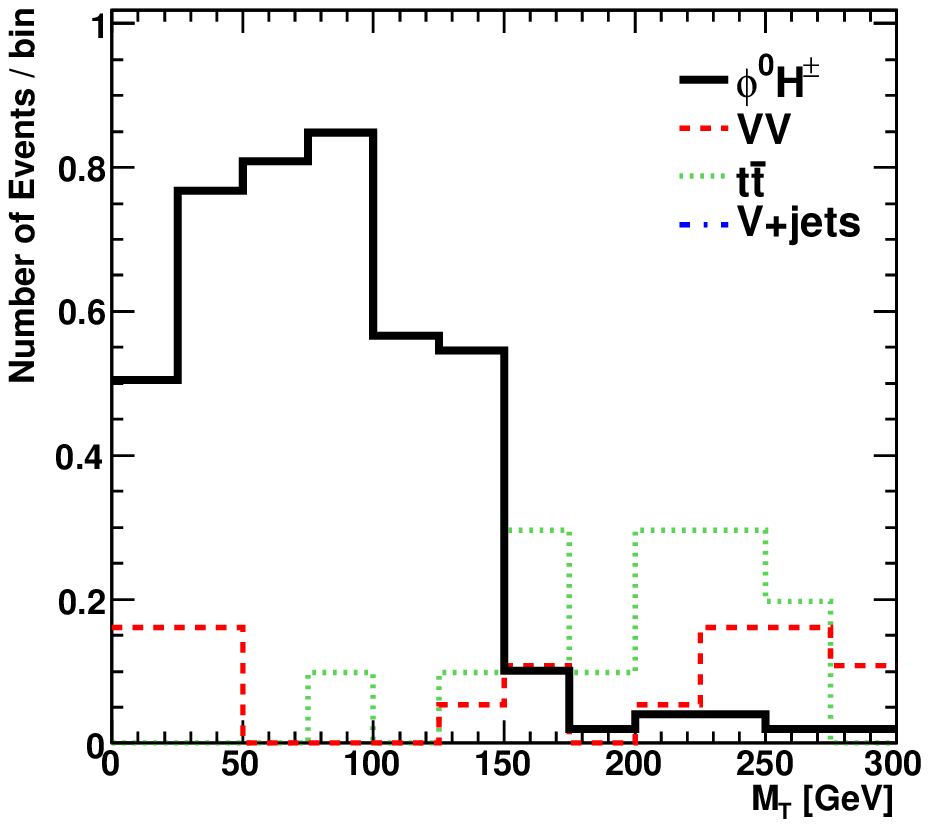}
 \includegraphics[height=7cm]{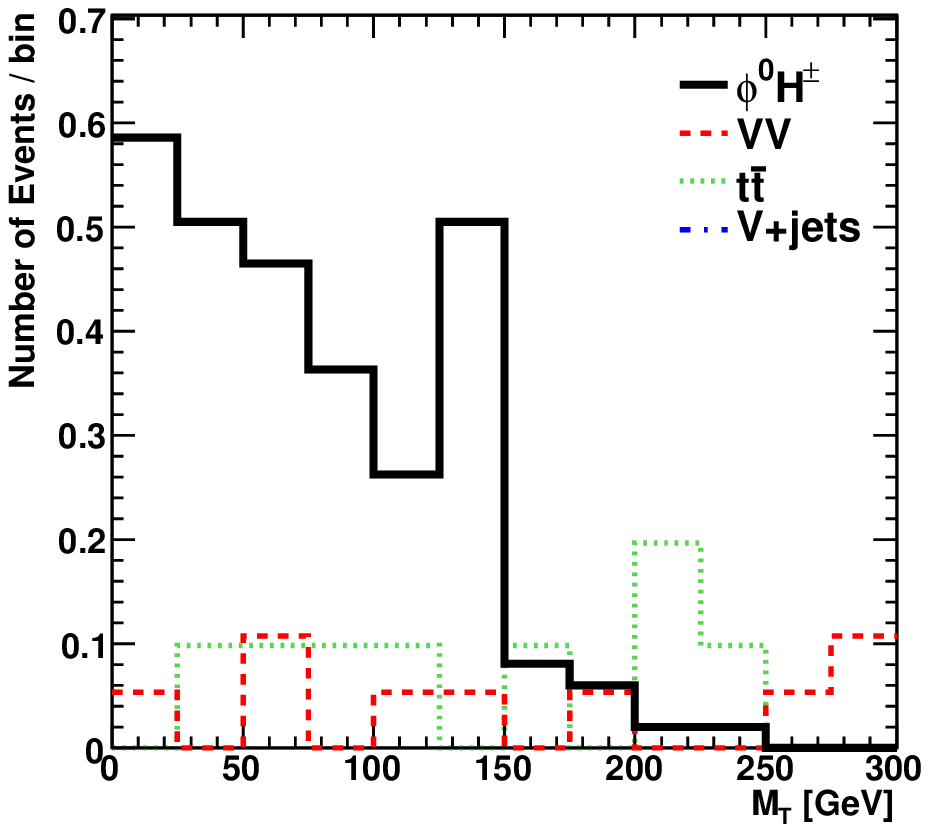}
 \caption{Distributions of the transverse mass of the tau-jet and the
 missing transverse momentum with
 $120~\text{GeV}<M_{\mu\mu}<140~\text{GeV}$ (left) and
 $160~\text{GeV}<M_{\mu\mu}<180~\text{GeV}$ (right) for the
 $2\mu1\tau_h$ channel after the selection cuts.
 The signal processes is simulated with $m_H^{}=130$ GeV, $m_A^{}=170$
 GeV and $m_{H^\pm}^{}=150$ GeV.}
 \label{FIG:MT_Mmm}
\end{figure}

In FIG.~\ref{FIG:MT}, two dimensional distributions of $M_{\mu\mu}\, 
\text{vs.}\, M_{T}$ are shown separately for the signal events (left)
and the background events (right). 
The signal distribution shows two bands in $0\lesssim M_T\lesssim
150$~GeV at $M_{\mu\mu}\simeq$ 130~GeV and 170~GeV.
Clearly, these bands indicate the signature of $\phi^0 H^{\pm}$ production
followed by $\phi^0\to\mu^+\mu^-$ and $H^\pm\to\tau^{\pm}\nu$ with the
hadronic decay of the tau lepton.
The transverse mass distribution in each band shows an edge at the mass
of the charged Higgs boson.
In FIG.~\ref{FIG:MT_Mmm}, we show the transverse mass distributions for
the mass window of $|M_{\mu\mu}-m_{\phi^0}^{}|\le 10$~GeV for $\phi^0=H$
(left) and $\phi^0=A$ (right).
Both distributions have an edge at $M_T\simeq 150$~GeV, which corresponds to 
the mass of the charged Higgs boson. 
However the expected number of events for $L=100$~fb$^{-1}$ is not large enough.
If a sufficient number of events is accumulated, 
it should be possible to measure the mass of the charged Higgs boson from 
the position of the edge in the transverse mass distribution around
$M_{\mu\mu}\simeq m_{H}^{}$ and $m_{A}^{}$.

\begin{table}[tb]
 \begin{center}
 \begin{tabular}{|c||cc|cc|cc|}
  \hline
  \smash{\lower7pt\hbox{Lepton channels}}
  & \multicolumn{2}{c|}{$2\mu1\tau_h$}
  & \multicolumn{2}{c|}{$3\mu$}
  & \multicolumn{2}{c|}{$2\mu1e$}
  \\ [.3mm]
  & $s/b$ & ($S$) & $s/b$ & ($S$) & $s/b$ & ($S$) \\
  \hline \hline
  Pre-selection
  & 445./2330550. & (0.3) & 34.1/4214. & (0.5) & 81.8/3880. & (1.3) \\
  $p_T^{\tau_h} > 40$~GeV
  & 301./465015. & (0.4) & - & (-) & - & (-) \\
  $\cancel{E}_T^{} > 80$~GeV
  & 135./3163. & (2.4) & 13.9/416. & (0.7) & 30.6/385. & (1.5) \\
  $H_T^\text{jet} < 30$~GeV
  & 73.1/320. & (3.9) & - & (-) & - & (-) \\
  $H_T^\text{lep} > 350$~GeV
  & 39.7/113. & (3.5) & 7.5/122. & (0.7) & 14.3/115.  & (1.3) \\
 $(m_Z^{})_{\mu\mu} \pm 10$~GeV
  & 36.2/74.4 & (3.9) & 7.0/67.8 & (0.8) & 13.2/11.9 & (3.3) \\
  \hline
  O.S.\ muons
  & 25.3/73.3 & (2.8) & - & (-) & 9.2/10.0 & (2.6) \\
  $p_T^{\mu\mu} > 80$~GeV
  & 17.1/41.5 & (2.5) & - & (-) & - & (-) \\
  $(m_{\phi^{0}}^{})_{\mu\mu} \pm 10$~GeV
  & 6.1/5.4 & (2.3) & 3.0/7.7 & (1.0) & 4.0/1.0 & (2.9) \\
  \hline
 \end{tabular}
 \end{center}
 \caption{Summary of event rejection by the selection cuts for the
 three-lepton channels with two or more muons.}
 \label{Tab:LHC_2MC}
\end{table}

Similar channels to the $2\mu1\tau_h$ channel are $3\mu$ and $2\mu1e$. 
We summarize the selection cuts for these process in
TABLE~\ref{Tab:LHC_2MC}.

\subsubsection{Three-lepton channels with two or more tau-jets}

Then, we study the three-lepton channels with two or more tau-jets.
Similarly to the four-lepton channels with two or more tau-jets, 
these channels are expected to have large number of events due to the
dominant branching ratios into tau leptons.
On the other hand, due to the many sources of the missing momentum,
namely neutrinos from the decay of the charged Higgs boson and the tau
leptons, the reconstruction of the event kinematics is limited. 

\begin{figure}[tb]
 \includegraphics[height=5.3cm]{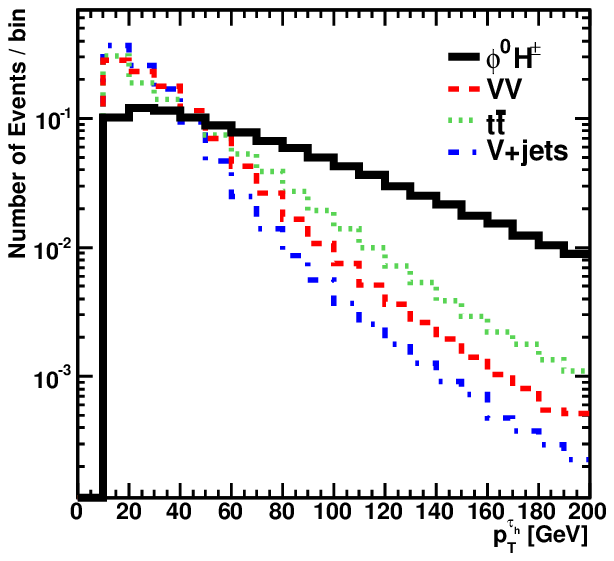}
 \includegraphics[height=5.3cm]{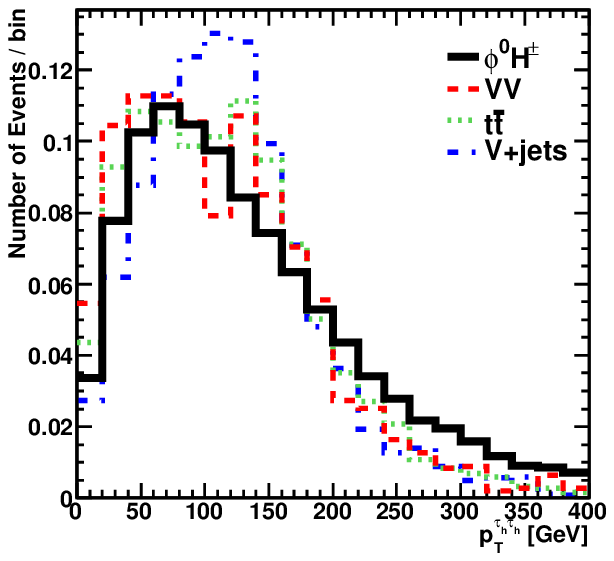} \hspace{5.3cm}\,
 \includegraphics[height=5.3cm]{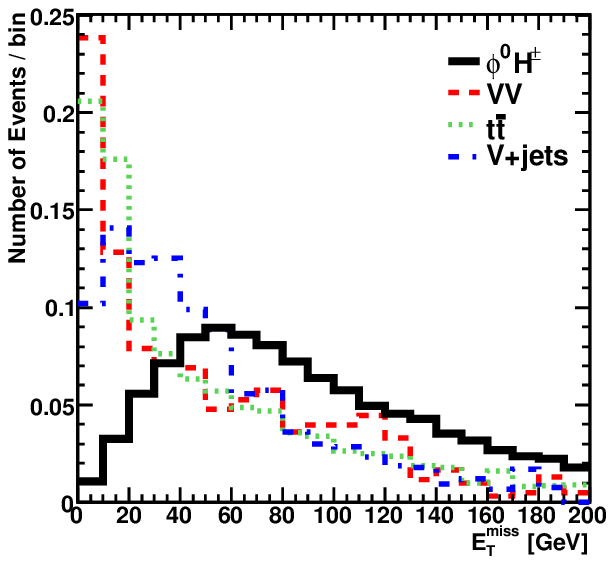}
 \includegraphics[height=5.3cm]{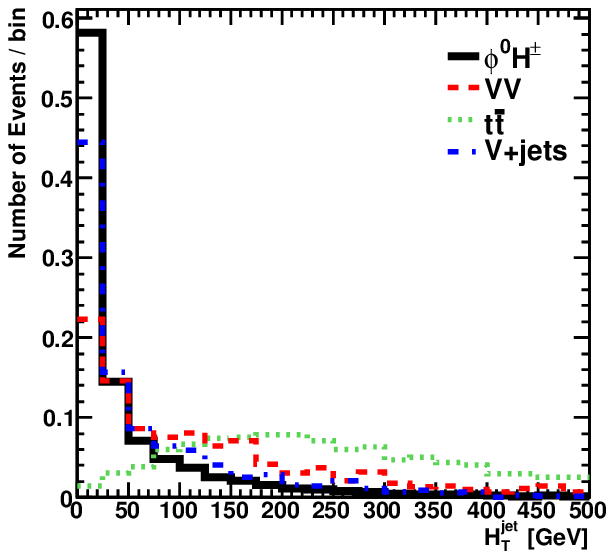}
 \includegraphics[height=5.3cm]{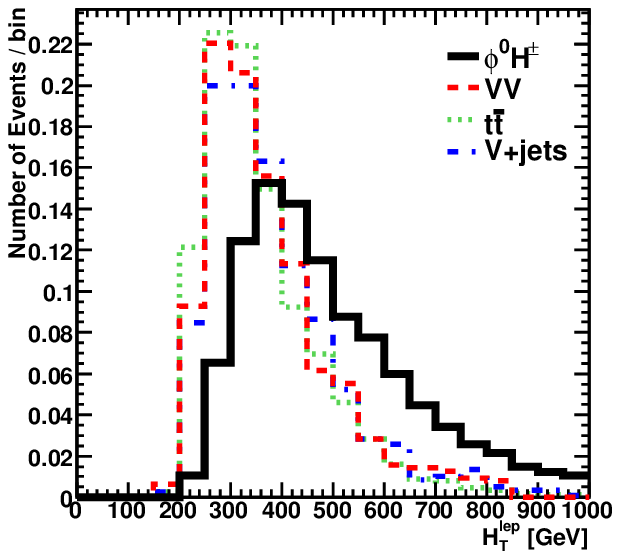}
 \caption{Distributions of kinematical variables for the $3\tau_h$ channel.
 Line descriptions follow those in FIG.~\ref{FIG:LHC_2T2M_Mmumu}.}
 \label{FIG:Dist_LHC_3T}
\end{figure}

Here, we present the analysis on the $3\tau_h$ channel and study
how the mass information can be extracted in this channel.
In FIG.~\ref{FIG:Dist_LHC_3T}, we show distributions of kinematical
variables, $p_T^{\tau_h}$, $p_T^{\tau_h\tau_h}$, $\cancel{E}_{T}$,
$H_T^\text{jet}$ and $H_T^\text{lep}$.
The numbers of the signal and background events are normalized to be unity. 
Taking into account these distributions, we apply the following
selection cuts for the signal and background events;
\begin{subequations}
\begin{align}
 & p_T^{\tau_h}    > 60~\text{GeV}, \\
 & \cancel{E}_T^{} > 80~\text{GeV}, \\
 & H_T^\text{jet}  < 50~\text{GeV}, \\
 & H_T^\text{lep}  > 350~\text{GeV}.
\end{align}
\end{subequations}
The summary of the results of event reductions is given in
TABLE~\ref{Tab:LHC_3TC}. 
After the all cuts, more than a hundred event is expected for the
signal processes, while the background events are estimated to be around
$120$ which is dominated by the $V+$jets process for $L=100$~fb$^{-1}$.
The expected reach of the significance is close to $10$ for 
$L=100$~fb$^{-1}$. 
It may indicate the possibility of an earlier evidence of the signal in
this channel. 
The larger $H_T^\text{lep}$ cut can be applied to enhance the
signal-to-background ratio, which however tends to reduce the
significance.

\begin{table}[tb]
\begin{center}
\begin{tabular}{|c||c|c||c|c|c||c|c|}
 \hline $3\tau_h$ event analysis & $HA$ & $\phi^0 H^{\pm}$ & $VV$ &
 $t\bar t$ & $V+$jets & $s/b$ & $S$ (100~fb$^{-1}$) \\ \hline \hline
 Pre-selection
 & 710.5 & 2476.  & 7069.  & 42053.  & 777043.  & $10^{-3}$ &  3.5 \\
 $p_T^{\tau_h} > 60$~GeV
 &  88.4 &  299.  &   34.2 &   341.  &   1237.  & 0.2 &  9.3 \\
 $\cancel{E}_T^{} > 80$~GeV
 &  29.3 &  165.  &    9.9 &    89.9 &    272.  & 0.5 &  9.3 \\
 $H_T^\text{jet} < 50$~GeV
 &  15.8 &  114.  &    4.0 &     4.8 &    123.  & 1.0 &  9.9 \\
 $H_T^\text{lep} > 350$~GeV
 &  15.0 &  106.  &    3.4 &     3.8 &    108.  & 1.0 &  9.9 \\
 \hline
\end{tabular}
\caption{Table for background reductions in the $3\tau_h$ channel.}
\label{Tab:LHC_3TC}
\end{center}
\end{table}

\begin{figure}[tb]
 \includegraphics[height=7cm]{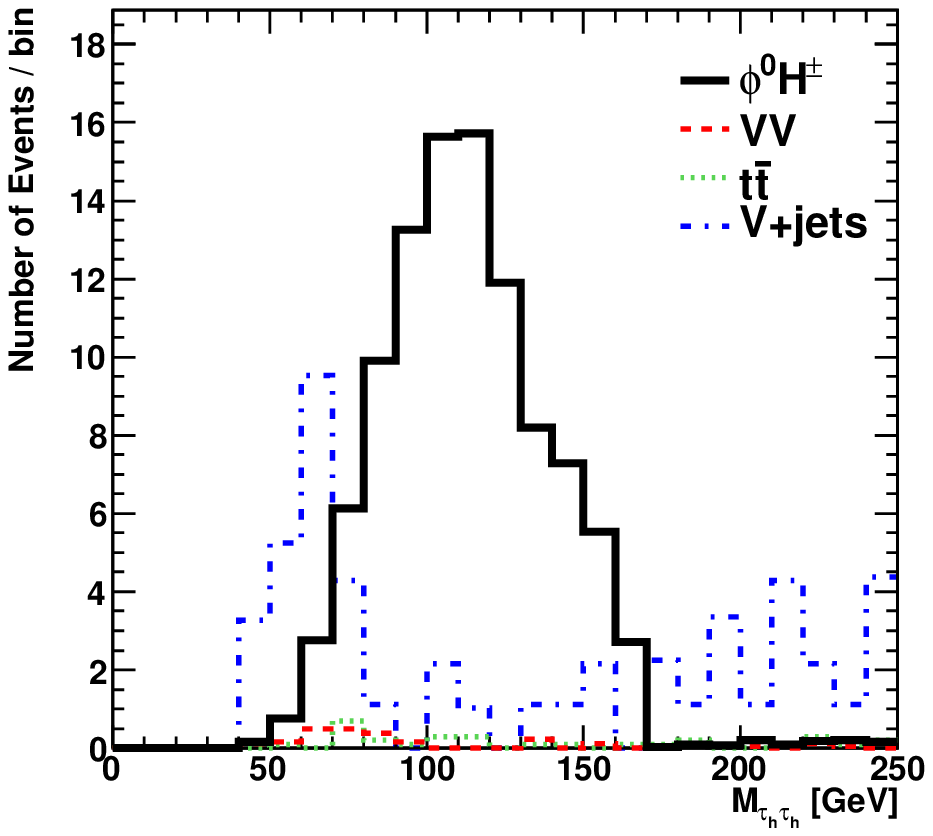}
 \includegraphics[height=7cm]{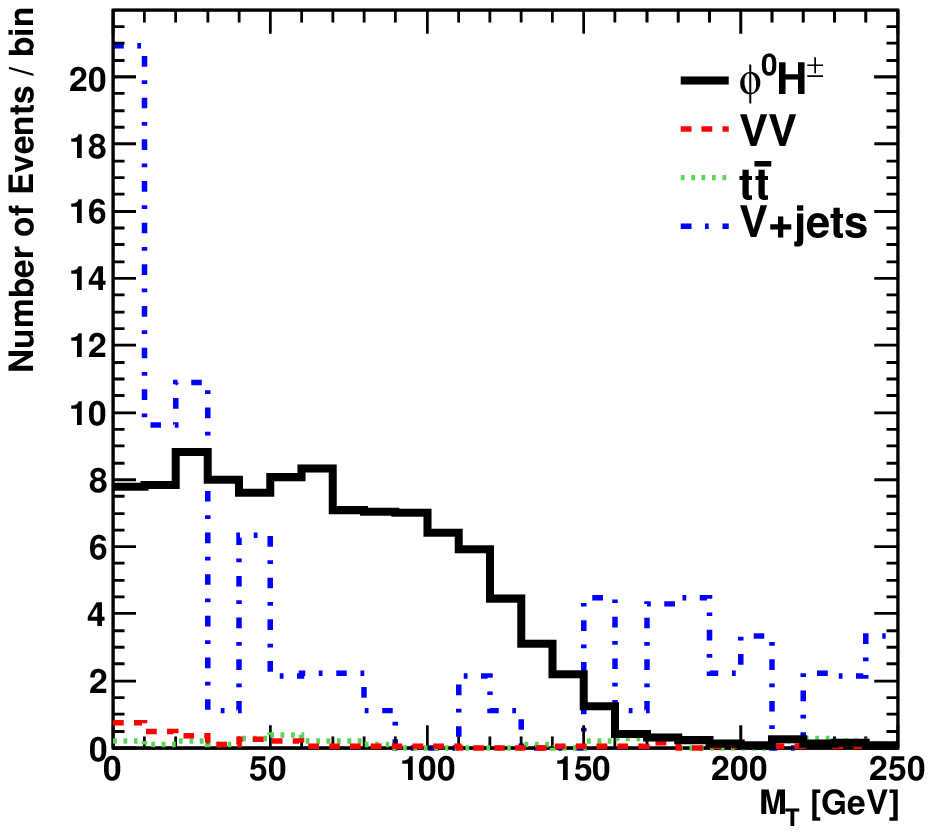}
 \caption{Distributions of the invariant mass of the tau-jet pair (left)
 and the transverse mass of the remaining tau-jet and the missing
 transverse momentum in the $3\tau_h$ channel after the selection cuts.
 The pairing rule of the tau-jets is explained in the text.}
 \label{FIG:LHC_3T_M}
\end{figure}

In the left panel in FIG.~\ref{FIG:LHC_3T_M}, we plot the invariant mass
distributions of the tau-jet pair with opposite charges after the
selection cuts for the signal and background processes, where the pair
of the two tau-jets is chosen such that it has opposite charges and
gives the largest transverse momentum $p_T^{\tau_h\tau_h}$. 
The $M_{\tau_h\tau_h}$ distribution shows an endpoint behavior at
$M_{\tau_h\tau_h}\simeq 170$~GeV which corresponds to the mass of the
heavier neutral Higgs boson.
Another sharp edge may be able to be pinned down at
$M_{\tau_h\tau_h}\simeq 130$~GeV the mass of the lighter neutral Higgs
boson.
In the right panel, distributions of the transverse mass $M_T$ 
constructed from the remaining tau-jet and the missing transverse
momentum using Eq.~\eqref{Eq:MT} are shown for the signal and background
events after the selection cuts in the $3\tau_h$ channel.
The $M_T$ distribution for the signal events vanishes around $M_T\simeq
150$~GeV, however the slope is, in contrast to the $2\mu1\tau_h$
channel, rather gentle but not seen like an endpoint.
This is because there are many sources of the missing momentum in this
channel.

\begin{figure}[tb]
 \includegraphics[height=7cm]{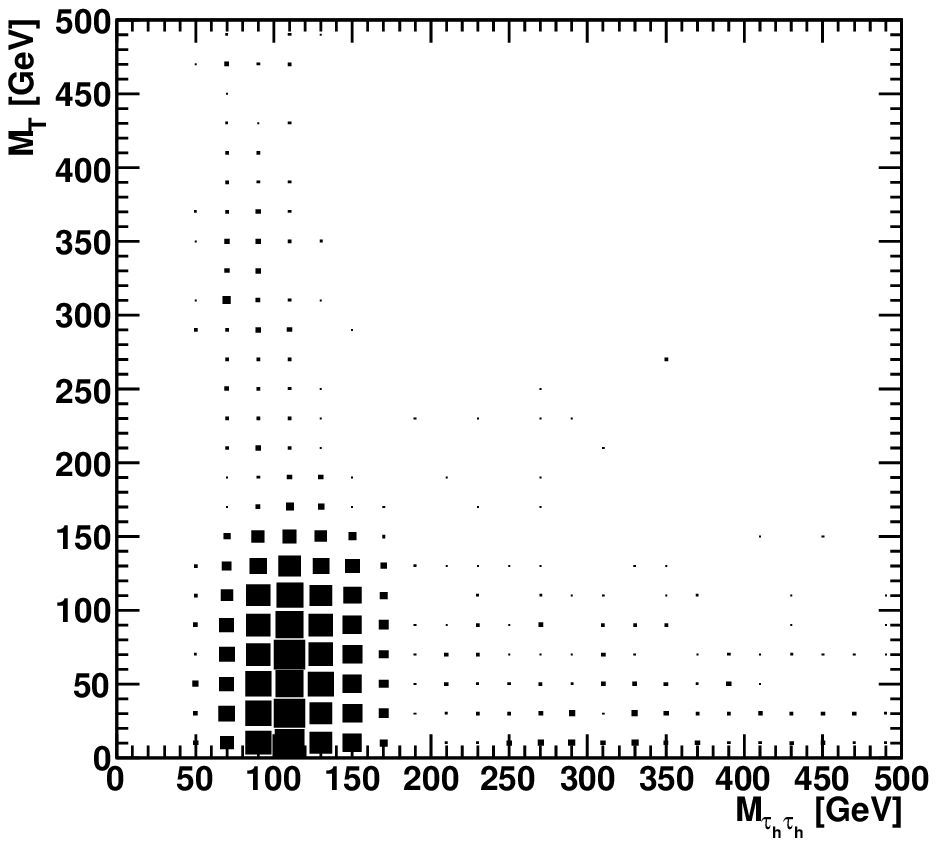}
 \includegraphics[height=7cm]{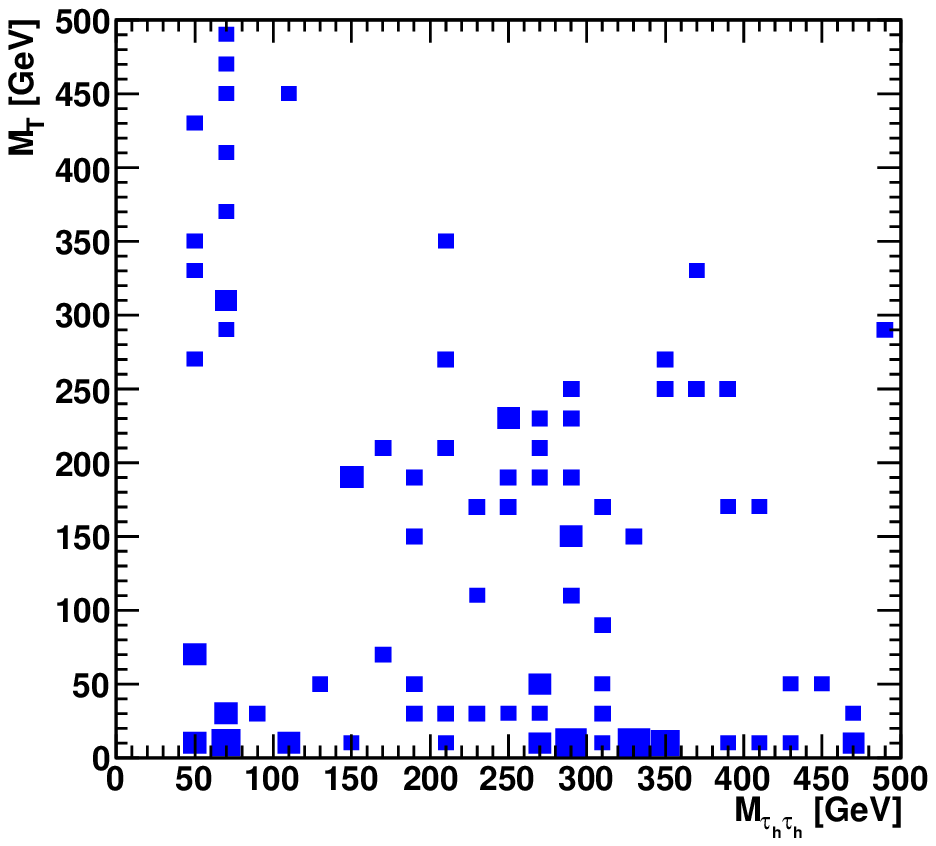}
 \caption{Two dimensional distributions of the invariant mass of the
 tau-jet pair versus the transverse mass in the $3\tau_h$ channel after
 the selection cuts.
 The left panel is for the signal process, and the right panel is for the
 background processes.
 The normalization of the two panels is adjusted.
 }
 \label{FIG:LHC_3T_M2}
\end{figure}

In the left panel in FIG.~\ref{FIG:LHC_3T_M2}, we plot the two dimensional
distribution of the invariant mass of the tau-jet pair and the
transverse mass in the $3\tau_h$ channel after the selection cuts.
In the right panel, the dominant $V+$jets background events are plotted.
The $V+$jets background events widely distribute in the two dimensional
space of $M_{\tau_h\tau_h}$ and $M_T$.
On the other hand, the signal events distribute compactly in the region
of $50~\text{GeV}\lesssim M_{\tau_h\tau_h}\lesssim\text{max}~(m_H,m_A)$
and $0\lesssim M_T\lesssim m_{H^{\pm}}^{}$.
Therefore, if the sufficient number of events is accumulated, we may
have an indication of the masses of the charged Higgs boson and the
neutral Higgs bosons by catching the border of the signal-event-like
distribution. \\


Similar channels to the $3\tau_h$ channel are $2\tau_h1\mu$ and
$2\tau_h1e$. 
We summarize the selection cuts for these processes in
TABLE~\ref{Tab:LHC_2TC}.
The three-lepton channels with one or less muon or electron suffer large
background contributions from the $W+$jets process, thus the $s/b$ ratio
is not so large.
However due to the large number of the signal events, the significance
$S$ can be large enough.
There are also $1\tau_h1\mu1e$, $1\tau_h2e$, $1\mu2e$ and $3e$ channels.
For the reference, we list the summary of reductions of events in
TABLE~\ref{Tab:LHC_1TC}.

\begin{table}[tb]
 \begin{center}
 \begin{tabular}{|c||cc|cc|cc|}
  \hline
  \smash{\lower7pt\hbox{Lepton channels}}
  & \multicolumn{2}{c|}{$3\tau_h$}
  & \multicolumn{2}{c|}{$2\tau_h1\mu$}
  & \multicolumn{2}{c|}{$2\tau_h1e$}
  \\ [.3mm]
  & $s/b$ & ($S$) & $s/b$ & ($S$) & $s/b$ & ($S$) \\
  \hline \hline
  Pre-selection
  & 3186./826165. & (3.5) & 2001./622055. & (2.5) & 1794./571334. &
  (2.4) \\
  $p_T^{\tau_h} > 60$~GeV
  & 387./1612. & (9.3) & 494./7346. & (5.7) & 443./7368. & (5.1) \\
  $\cancel{E}_T^{} > 80$~GeV
  & 194./372. & (9.3) & 244./1940. & (5.4) & 216./1805. & (5.0) \\
  $H_T^\text{jet} < 50$~GeV
  & 129./132. & (9.9) & 158./761. & (5.5) & 142./738. & (5.1) \\
  $H_T^\text{lep} > 350$~GeV
  & 121./116. & (9.9) & 124./434. & (5.7) & 112./437. & (5.2) \\
  \hline
 \end{tabular}
 \end{center}
 \caption{Summary of event rejection by the selection cuts for the
  three-lepton channels with two or more tau-jets.}
 \label{Tab:LHC_2TC}
\end{table}
\begin{table}[tb]
 \begin{center}
 \begin{tabular}{|c||cc|cc|cc|cc|cc|}
  \hline
  \smash{\lower7pt\hbox{Lepton channels}}
  & \multicolumn{2}{c|}{$1\tau_h1\mu1e$}
  & \multicolumn{2}{c|}{$1\tau_h2e$}
  & \multicolumn{2}{c|}{$1\mu2e$}
  & \multicolumn{2}{c|}{$3e$}
  \\ [.3mm]
  & $s/b$ & ($S$) & $s/b$ & ($S$) & $s/b$ & ($S$) & $s/b$ & ($S$) \\
  \hline \hline
  Pre-selection
  & 738./65150. & (2.9) & 333./1883890. & (0.2) & 68.8/3431. & (1.2) &
  20.0/3107. & (0.4) \\
  $p_T^{\tau_h} > 40$~GeV
  & 498./17718. & (3.7) & 226./373858. & (0.4) & - & (-) & - & (-) \\
  $\cancel{E}_T^{} > 80$~GeV
  & 302./4971. & (4.2) & 97.3/2456. & (2.0) & 22.7/343. & (1.2) &
  6.7/302. & (0.4) \\
  $H_T^\text{jet} < 30$~GeV
  & 113./455. & (5.1) & 52.8/259. & (3.2) & - & (-) &
  - & (-) \\
  $H_T^\text{lep} > 350$~GeV
  & 58.5/128. & (4.8) & 27.7/94.4 & (2.7) & 9.1/99.5 & (0.9) & 2.7/90.7
  & (0.3) \\
  $(m_Z)_{ee}\pm 10$~GeV
  & - & (-) & 24.2/62.8 & (2.9) & 7.9/11.2& (2.2) & 2.2/39.9 & (0.3) \\
  \hline
 \end{tabular}
 \end{center}
 \caption{Summary of event rejection by the selection cuts for the
  rest of three-lepton channels.}
 \label{Tab:LHC_1TC}
\end{table}
%

\section{Summary and Discussions}

In this section, we present a summary and discussions of our simulation
studies, and give discussions for the neutral Higgs pair production at
electron-positron linear colliders. 
Finally, we conclude the paper. 

We have studied multi-$\tau$ signatures at the LHC in the Type-X THDM. 
Assuming $\tan\beta \gtrsim 3$, more than $99$\% of extra Higgs bosons 
$H,A$ and $H^\pm$ decays into tau leptons. 
These Higgs bosons can be pair produced in $q\bar q\to HA$ and $q\bar
q'\to \phi^0 H^\pm$ processes where $\phi^0=H$, $A$, and they produce 
characteristic multi-$\tau$ final states at the LHC. 
We have focused on the $2\mu2\tau$ and $4\tau$ states for the pair
production of the neutral Higgs bosons, and on the $2\mu1\tau1\nu$ and 
$3\tau1\nu$ states for the charged Higgs boson associated production
followed by the decays of tau leptons into leptons or hadrons. 
These result in the various four- and three-lepton channels 
which include tau-jets.
%
We have performed the Monte-Carlo simulation for the signal processes
as well as the major background processes, namely $VV$, $t\bar t$ and
$V+$jets processes at the LHC where $V=W$, $Z$.
The realistic tau-jet tagging procedure is simulated for the final state
hadrons.
For leptons, the standard acceptance cuts and isolation conditions
are considered.
Then we have applied the selection cuts to reduce the background
contributions for each channel.
We here summarize the analysis for the four- and three-lepton channels
to probe the pair production of the neutral Higgs bosons and the
charged Higgs boson associated production, respectively, in order.
\begin{itemize}
\item
     For the four-lepton channels with two or more muons, 
     the mass of the neutral Higgs bosons can be measured as 
     the peaks in the $M_{\mu\mu}$ distribution.
     For these events where the dimuon comes from the direct decay of
     the Higgs boson, the tau lepton momenta can be reconstructable by 
     using the collinear approximation for the rest two leptons out of
     the four. 
     The two dimensional distribution of the $M_{\mu\mu}$ and the
     reconstructed $M_{\tau\tau}$, such like shown in
     FIG.~\ref{FIG:LHC_2T2M_Mmumu_Mtjtj}, 
     would be a direct test of the neutral Higgs bosons pair production
     in the Type-X THDM. 
     However, the numbers of the signal events are relatively small due
     to the small branching ratio of the neutral Higgs bosons into
     dimuons.
     The signal-to-background ratio is rather large, thus a clean
     extraction is possible, if the number of events is sufficient.
     The information of the branching ratio of the neutral Higgs boson
     into dimuons can also be extracted. 
     
     The detailed analysis is performed for the $2\mu2\tau_h$ channel,
     and the summary of the selection cuts for the similar channels is
     found in TABLEs~\ref{Tab:LHC_2T2M} and \ref{Tab:LHC_2M}. 
     
\item
     For the four-lepton channels with two or more tau-jets, a large
     number of signal events are expected due to the dominant branching
     ratio of the neutral Higgs bosons into tau leptons.
     After the selection cuts, both the signal-to-background ratios
     $s/b$ and the significances $S$ are large, so that the excess can
     be relatively easily found.
     On the other hand, the mass determination would only be possible
     by looking at the kinematical distributions. 
     Searches for the endpoints or bumps in the $M_{\tau_h\tau_h}$
     distributions, shown in FIGs.~\ref{FIG:LHC_4T_Mtjtj} and
     \ref{FIG:LHC_4T_Mtjtj1_Mtjtj2} can indicate the mass of the neutral
     Higgs bosons, provided the sufficient number of events are
     accumulated.
     
     The analysis for the $4\tau_h$ channel is presented in detail, and
     the selection cuts for the  similar channels are summarized in
     TABLEs~\ref{Tab:LHC_4T} and \ref{Tab:LHC_2T}.
     For completeness, a summary of the selection cuts for the rest of
     the four-lepton channels is presented in TABLE~\ref{Tab:LHC_1T}.
     
\item
     For the three-lepton channels with two or more muons, 
     the mass of the neutral Higgs bosons can be measured by looking
     for the peaks in the $M_{\mu\mu}$ distribution. 
     The mass of the charged Higgs boson can be measured as the edge of 
     the distributions of the transverse mass $M_T$ 
     which is constructed from the rest of the
     three-leptons and the missing transverse momentum.
     By focusing on the events with $M_{\mu\mu}\simeq m_{H}^{}$ or
     $m_{A}^{}$, an edge can be enhanced such
     as shown in FIG.~\ref{FIG:MT_Mmm}, which corresponds to the
     charged Higgs boson mass.
     Due to the large background contribution from the $V+$jets process,
     the $s/b$ ratio is small.
     
     The analysis for the $2\mu1\tau_h$ channel is presented in detail,
     and the selection cuts for the the similar channels are
     summarized in TABLEs~\ref{Tab:LHC_1T2M} and \ref{Tab:LHC_2MC}.
     
\item
     For the three-lepton channels with two or more tau-jets, a large
     number of signal events is expected due to the dominant branching
     ratio of Higgs bosons into tau leptons. 
     However due to the large background contribution from the $V+$jets
     process, the $s/b$ ratio is rather small even after the selection
     cuts.
     Using the two dimensional distributions of $M_{\tau_h\tau_h}$ and
     the transverse mass $M_T$, as shown in FIG.~\ref{FIG:LHC_3T_M2},
     the information of the mass of the Higgs bosons can be obtained.
     
     The analysis for the $3\tau_h$ channel is presented in detail, and
     the selection cuts for the  similar channels are summarized in
     TABLEs~\ref{Tab:LHC_3TC} and \ref{Tab:LHC_2TC}.
     For completeness, a summary of the selection cuts for the rest of
     the three-lepton channels is presented in TABLE~\ref{Tab:LHC_1TC}.
     
\end{itemize}

We have shown that the characteristic excess from the production and 
the decay of the lepton-specific Higgs
bosons can be observed in the various four- and three-lepton channels on
top of the SM contributions at the LHC. 
However, on the other hand, the determination of the mass of the Higgs
bosons seems not to be feasible with the integrated luminosity of 
$L=100$~fb$^{-1}$, although we have shown the various methods in various
channels. 
The measurement of the branching ratio of the neutral Higgs bosons into
dimuons which we have examined in the $2\mu2\tau_h$ channel, also
requires the number of events more than the expected for
$L=100$~fb$^{-1}$. 
These measurement should be achieved with the running of the luminosity
upgraded LHC in future~\cite{Ref:SLHC}. 
With the luminosity of an order of thousand inverse fb, it is promising
to measure the mass of the neutral and charged Higgs bosons as well as
the branching ratio of the neutral Higgs boson into dimuons. \\

We here briefly comment on the search potential for the lepton-specific
Higgs bosons at the International Linear Collider (ILC) or the Compact
Linear Collider (CLIC).
The neutral Higgs bosons can be pair produced via the $e^+e^-\to HA$
process, and their decay produces four-$\tau$ states predominantly. 
Through the leptonic and hadronic decay of the tau leptons, the
signatures are four-lepton channels including tau-jets, as we have
studied for the hadron colliders.
At $e^+e^-$ colliders, four momenta of the four tau leptons can be
solved by applying the collinear approximation to all the four decay
products of the tau leptons~\cite{Ref:LEP4tau,Ref:LEPdouble}, because
the missing four momentum can be reconstructed by the energy momentum
conservation. 
In Ref.~\cite{Ref:4tauILC}, it is shown that clear peaks can be observed
in the one dimensional and two dimensional distributions for the
invariant masses of the reconstructed tau lepton pairs. \\

In conclusion, we have studied the multi-$\tau$ signatures at the LHC
in the THDM with lepton-specific Yukawa interactions.
At the LHC, such Higgs bosons can be pair produced via the weak boson
mediated processes, and predominantly they decay into multi-$\tau$
states. 
Although the event reconstruction is limited due to the missing energies
from the decays of the tau leptons, it is possible to discriminate the
signal events from the background events by requiring the high
multiplicity of the leptons and the tau-jets with appropriate
kinematical cuts in various four and three-lepton channels.
The mass of the Higgs bosons can be measured by using the invariant
mass distributions of the dimuon which comes from the direct decay of
the neutral Higgs bosons as well as the endpoint analysis for the
distributions of the invariant mass of the tau-jets and the transverse
mass, if sufficiently large integrated luminosity is accumulated.
We have found that measurements of multi-$\tau$ final states at hadron
colliders can be a useful probe of the properties of the Higgs bosons in
the lepton-specific models.

\acknowledgments
We thank S.~Odaka and the KEK Atlas group for providing us and managing
computing environment to execute the Monte-Carlo simulation in this
work.
The work of S.K.\ was supported in part by Grant-in-Aid for Scientific
Research, Japan Society for the Promotion of Science (JSPS),
Nos.~22244031 and 23104006.
The work of K.T.\ was supported in part by the National Science Council
of Taiwan under Grant No.~NSC 100-2811-M-002-090.
The work of H.Y.\ was supported in part by the National Science Council
of Taiwan under Grant No.~NSC 100-2119-M-002-001.
K.T.\ and H.Y.\ would like to thank the theoretical physics group in
university of Toyama for their warm hospitality.

\appendix
\section{Tau-jet tagging efficiency}\label{app:tag}

In this appendix, we present a rough estimation of the tau-jet tagging
efficiency and the mis-tagging probabilities for the non-tau jet
in our tau-jet tagging procedure defined in the main text.
We neglect the $\eta$ and $p_T^{}$ dependences of the efficiency
as well as the mis-tagging probability.
This information may be useful to estimate the variation of our results
by changing the conditions of tau-jet tagging, and also to compare with
the actual experimental situation at the LHC. 

The tau-jet tagging efficiency is estimated from the signal events
in $4\tau_h$ and $3\tau_h$ channels from $HA$ production and
$\phi^{0}H^{\pm}$ production, respectively.
In the $4\tau_h$ channel, the number of events after the pre-selection
for the signal events is $\sim320$ for the integrated luminosity of
100~fb$^{-1}$ (see TABLE~\ref{Tab:LHC_4T}).
On the other hand, the expected number of the $4\tau$ final-state is
$50~[\text{fb}]\times100~[\text{fb}^{-1}]\sim5000$.
From this, the probability of finding the tau-jet per decay of one tau
lepton is estimated to be $(320/5000)^{1/4}\sim0.50$.
This can be interpreted as the product of the branching ratio of the
hadronic decay of tau leptons, the acceptance ratio for $|\eta|<2.5$ and
$p_{T}^{}>10$~GeV and also the tau-jet tagging efficiency.
By using the branching ratio of hadronic decays of tau leptons which
is about 65\%~\cite{Ref:PDG}, the multiplication of the
the acceptance ratio and tau-jet tagging efficiency is estimated to be
$\sim 0.77$.
Further separation of the two factors is too confusing to us.
For confirmation we have checked that the obtained value gives roughly
correct number of the signal events from $\phi^0H^{\pm}$ production in
the $3\tau_h$ channel by
$200~[\text{fb}]\times100~[\text{fb}^{-1}]\times(0.50)^3\sim2500$
(compare with the number of signal events from $\phi^{0}H^{\pm}$
production after the pre-selection in TABLE~\ref{Tab:LHC_3TC}).

The mis-tagging probability for non-tau jets is estimated for the
$V+$jets events in the $2\mu2\tau_h$ and $2\mu1\tau_h$ channels.
The numbers of events from the $V+$jets production after the
pre-selection are 2.9$\cdot 10^4$ and 2.3$\cdot 10^6$ for the
$2\mu2\tau_h$ and $2\mu1\tau_h$ channels, respectively, where here the 
requirement of the zero charge-sum is not imposed in the former
channel.
For our $V+$jets events, the fraction of the events which exclusively
contain one (two) primal jet(s) with $|\eta|<2.5$ and $p_{T}^{}>10$~GeV
is found to be 0.09 (0.01).
The expected number of events for the $\mu^{+}\mu^{-}jj$ final-state
from the $V+$jets process is $30~[\text{nb}]\times
100~[\text{fb}^{-1}]\times0.33\times0.01\sim10^{7}$,
where a factor of 0.33 comes from the fraction of the dimuon final-state
in the leptonic $Z$-boson decays.
Thus the mis-tagging probability per jet is estimated to be
$\sqrt{2.9\cdot10^4/10^7}\sim 0.05$.
The same probability is estimated for the $2\mu1\tau_h$ channel as
$2.3\cdot10^6/(10^9\times0.09)\sim 0.03$.
Thus the mis-tagging probability per jet for the $V+$jets process should 
be a few percent in our simulation.

Similarly the mis-tagging probability for the $t\bar t$ events can be
estimated by using the dimuon decay of the $t\bar t$ events.
For the $t\bar t$ events, one or two jets are expected to be originated
from the $b$ quark, and the mis-tagging probability for such jets may be
different from that for the light-flavor jets.
However, we ignore the difference of the mis-tagging probabilities by
the flavor of jets.
The number of events for the dimuon decay mode of our $t\bar t$ events
is given as
$493~[\text{pb}]\times100~[\text{fb}^{-1}]\times (0.107)^2\sim
5.6\cdot10^{5}$.
For the dimuon decay channel in our $t\bar t$ events, an average number
of the primal jet with $|\eta|<2.5$ and $p_{T}^{}>10$~GeV is
$\sum_{n=1}r_n\,n\sim3.6$ where $r_n$ is the fraction of events with $n$
primal-jets which satisfies $\sum_{n=0}r_n=1$, and also an average
number of the combination of the two primal jets is $\sum_{n=2}r_n\,{n
\choose 2}\sim5.8$.
By removing the requirement of zero charge-sum of the four-lepton at
the pre-selection in the $2\mu2\tau_h$ channel, the number of events
after the pre-selection is $\sim 800$, thus the mis-tagging probability
is estimated to be $\sqrt{800/(5.6\cdot10^{5}\times5.8)}\sim 0.02$.
On the other hand, for the $2\mu1\tau_h$ channel, the mis-tagging
probability is estimated to be $2.3\cdot10^4/(5.6\cdot10^5\times
3.6)\sim0.01$.
Thus the mis-tagging probability per jet for the $t\bar t$ process
should be also a few percent.
The smaller mis-tagging probability for the $t\bar t$ process by our
rough estimation may be due to the large QCD activity in the $t\bar t$
events which could prevent the isolation of the charged tracks in the
tau-jets.

One may be able to reduce the mis-tagging probability with minor loss
of tagging efficiency, e.g., by using only the 1-prong (or 3-prong) 
decay mode of tau leptons and/or by applying more tight tagging
conditions.


\end{document}